\documentclass[a4paper,twocolumn,aps,pre]{revtex4-1}

\usepackage{mathpazo}

\usepackage[T1]{fontenc}
\usepackage[latin9]{inputenc}
\usepackage{color}
\usepackage{mathrsfs}
\usepackage{amsmath}
\usepackage{amssymb}
\usepackage{graphicx}
\usepackage{wasysym}
\usepackage[colorlinks,citecolor=blue,filecolor=black,linkcolor=red,urlcolor=blue]{hyperref}
\usepackage{breakurl}

\makeatletter

\usepackage{lipsum}
\usepackage{tikz}
\makeatother

\begin{document}

\title{Scattering of solitons in binary Bose-Einstein condensates \\with
spin-orbit and Rabi couplings}

\author{Rafael M. P. Teixeira}
\affiliation{Instituto de F\'isica, Universidade Federal de Goi\'as, 74.690-900, Goi\^ania,
Goi\'as, Brazil}

\author{Wesley B. Cardoso}
\email{wesleybcardoso@gmail.com}
\affiliation{Instituto de F\'isica, Universidade Federal de Goi\'as, 74.690-900, Goi\^ania,
Goi\'as, Brazil}

\begin{abstract}
In this paper we study the scattering of solitons in a binary Bose-Einstein
Condensate (BEC) including SO- and Rabi-couplings. To this end, we
derive a reduced ODE model in view to provide a variational description
of the collisional dynamics. Also, we assume negative intra- and inter-component
interaction strengths, such that one obtains localized solutions even
in absence of external potentials. By performing extensive numerical
simulations of this model we observe that, for specific conditions,
the final propagation velocity of the scattered solitons could be
highly sensitive to small changes in the initial conditions, being
a possible signature of chaos. Additionally, there are infinitely
many intervals of regularity emerging from the obtained chaotic-like
regions and forming a fractal-like structure of reflection/transmission
windows. Finally, we investigate how the value of the spin-orbit coupling
strength changes the critical velocities, which are minimum/maximum
values for the occurrence of solitons bound-states, as well as the
fractal-like structure.
\end{abstract}

\maketitle

\section{Introduction \label{sec:I}}

Spin-orbit (SO) coupling was recently engineered in a neutral atomic
Bose-Einstein condensate (BECs) by dressing two atomic spin states
(hyperfine states $|F=1,\,m_{F}=\pm1\rangle$ of a spin-$1$ $^{87}\text{Rb}$
BEC) with a pair of laser beams \cite{Lin_NAT11}. This new scenario
has motivated further studies on vector solitons and other nonlinear
waves, such as, self-trapped states \cite{Merkl_PRL10}, vortices
\cite{Xu_PRL11,Radic_PRA11,Ramachandhran_PRA12,Lobanov_PRL14,Sakaguchi_PRE14-2},
Skyrmions \cite{Kawakami_PRL12}, Dirac monopoles \cite{Conduit_PRA12},
dark solitons \cite{Fialko_PRA12,Achilleos_EPL13}, bright solitons
\cite{Achilleos_PRL13}, gap solitons \cite{Kartashov_PRL13,Zhang_PRA15,Sakaguchi_PRA18},
exotic complexes \cite{Belobo_SR18}, etc. Furthermore, many studies
in BECs with SO coupling have shown interesting effects like the chiral
confinement in quasirelativistic BECs \cite{Merkl_PRL10}, existence
of a \textquoteleft stripe phase\textquoteright{} \cite{Ho_PRL11,Sinha_PRL11},
tunneling dynamics \cite{Zhang_PRA12,Garcia-March_PRA14,Wang_PRA15},
the partial wave scattering \cite{Williams_SCI12}, the phenomenon
of Zitterbewegung \cite{LeBlanc_NJP13,Qu_PRA13,Achilleos_PRA14},
the tunability of the SO coupling strength \cite{Jimenez-Garcia_PRL15},
traveling Majorana solitons \cite{Zou_PRL16}, steadily moving solitons
in a helicoidal gauge potential \cite{Kartashov_PRL17}, negative-mass
hydrodynamics \cite{Khamehchi_PRL17}, etc.

Analytical developments for search localized solutions in BECs with
SO coupling was recently reported in quasi-one- \cite{Salasnich_PRA13,Kartashov_PRL13,Achilleos_PRL13,Xu_PRA13,Zezyulin_PRA13,Achilleos_EPL13,Kartashov_PRA14,Sakaguchi_PRE14,Chiquillo_LP14,Gautam_PRA15,Zhang_PRA15,Cao_JOSAB15,Wen_PRA16,Li_CPL16,Sakaguchi_PRA17,Li_NJP17,Belobo_SR18,Sakaguchi_PRA18}
and quasi-two-dimensional \cite{Salasnich_PRA14,Sakaguchi_PRE14,Sakaguchi_PRE14-2,Liao_PRA17,Li_PRA17,Kato_PRA17,Sakaguchi_PRA18,Huang_PRA18}
systems. Specifically, in Ref. \cite{Salasnich_PRA13} was derived
an effective 1D coupled nonpolynomial Schr\"odinger equations from the
system of 3D Gross-Pitaevskii equations. Next, this study was extended
to quasi-two-dimensional BECs with SO and Rabi couplings \cite{Salasnich_PRA14}.
Detailed studies of stationary and moving bright solitons in BECs
with SO and Rabi couplings was presented in Refs. \cite{Xu_PRA13,Liu_EPL14,Li_CPL16,Wen_PRA16,Sakaguchi_PRA17,Li_NJP17}
and in Refs. \cite{Chiquillo_LP14,Liao_PRA17} including also interatomic
magnetic dipole-dipole interactions. In Ref. \cite{Zezyulin_PRA13}
was reported the existence of even, odd, and asymmetric nonlinear
modes in the effectively 1D self-repulsive binary BEC with the SO
and Zeeman splitting, confined by the axial HO potential. The emergence
of a number of nontrivial soliton properties due to a localized SO
coupling was presented in Ref. \cite{Kartashov_PRA14}. In Ref. \cite{Sakaguchi_PRE14}
was studied discrete and continuum composite solitons in BECs with
the Rashba SO coupling loaded into a deep 1D or 2D optical-lattice
potential. The spontaneous symmetry breaking in a SO-coupled $f=2$
spinor condensate was reported in \cite{Gautam_PRA15}. In Ref. \cite{Cao_JOSAB15}
was numerically investigated the ground state properties and dynamical
generation of dark solitons in SO-coupled BECs. Recently, was reported
in Ref. \cite{Huang_PRA18} the possibility to stabilize excited states
of semi-vortex and mixed-mode solitons (originally unstable) in a
setting based on repulsive dipole-dipole interactions induced by a
polarizing field, oriented perpendicular to the plane in which the
dipolar BEC is trapped. In addition, it has also been predicted that
2D and 3D solitons can be stabilized in spinor (two-component) BECs
with the help of Rashba-type SO coupling \cite{Wilson_PRL13,Sakaguchi_PRE14-2,Jiang_PRA16,Sakaguchi_PRE14,Gautam_PRA17,Chen_CNSNS17,Zhang_PRL15,Li_NJP17,Liao_PRA17}.

In a more complex scenario, collisions of solitary waves can show
nontrivial structures since, due to the nonintegrability of the system,
the collision outcome can depend on the initial conditions, presenting
in some cases a fractal pattern \cite{Yang_PRL00,Tan_PRE01,Dmitriev_CHAOS02,Dmitriev_PRE02,Zhu_PRE07,Zhu_PRL08,Zhu_PD08,Zhu_SAM09,Hause_PRA10,Teixeira_PLA16}.
Fractal structures in collisions of solitons are also reported in
systems described by other models, such as, in the $\phi^{4}$ model
\cite{Goodman_CHAOS08,Goodman_CHAOS15}, the sine-Gordon model \cite{Fukushima_PLA95,Higuchi_CSF98,Dmitriev_PRE01,Dmitriev_PB02,Dmitriev_PRE08},
etc. However, there are still few works dedicated to exploring collisions
of localized structures in BECs with SO coupling \cite{Sakaguchi_PRE14,Sakaguchi_PRE14-2,Kartashov_PRL17,Li_PRA17,Gautam_PRA17}.
Indeed, in Ref. \cite{Kartashov_PRL17} was reported the existence
and stability of families of steadily moving solitons in a helicoidal
gauge potential, where in the absence of Zeeman splitting, such solitons
interact elastically similarly to solitons in integrable systems.
Also, in Ref. \cite{Sakaguchi_PRE14-2} was verified that in two-dimensional
SO-coupled self-attractive BECs in free space, collisions between
two moving solitons lead to their merger into a single one. The scattering
process due to the collisions of solitons was used in Ref. \cite{Sakaguchi_PRE14}
in view to verify the stability of 1D and 2D solitons. In Ref. \cite{Li_PRA17}
it was studied the mobility and collision of gap-solitons in dipolar
BECs with SO coupling, revealing negative and positive effective masses
of the isotropic and anisotropic solitons, respectively. In addition,
in Ref. \cite{Gautam_PRA17} it was presented the study of the formation
and dynamics of 2D vortex-bright solitons in a three-component SO
coupled spin 1 spinor condensate, revealing that in the collision
of two moving vortex-bright solitons at small velocities, one finds
that the in-phase solitons either collapse or merge into a single
entity, whereas out-of-phase solitons repel and avoid each other without
ever having an overlapping profile. Here, we investigate the influence
of the SO coupling on the collisional dynamics of solitons in BECs.
To this end, we employ a reduced ordinary differential equations (ODE)
model based on a variational approach, which allow us to analytically
investigate the formation of fractal-like patterns and the properties
of the scattered solitons.

The rest of the paper is organized as follows. In Sec. \ref{sec:II},
we describe the effective mean-field coupled Gross-Pitaevskii (GP)
equations with SOC used to study the collisional dynamics of solitons.
By means of a variational approach, we obtain a reduced ODE model
in Sec. \ref{sec:III}. In Sec. \ref{sec:IV} we analyze the width
oscillations in the $|\xi|\gg1$ regime and the initial conditions
to be used in the numerical simulations presented in Sec. \ref{sec:V}.
Finally, in Sec. \ref{sec:Conclusion}, we give a summary of our findings.

\section{Theoretical Model \label{sec:II}}

We start by considering a BEC confined in a quasi-one-dimensional
parabolic trap (with frequencies $\omega_{x}\ll\omega_{\perp}$),
described by an effective 1D-GP equation system with SO and Rabi couplings,
which is written in a scaled form as \cite{Achilleos_PRL13} (length
in units of $a_{\perp}\equiv\sqrt{\hbar/m\omega_{\perp}}$, time in
units of $\omega_{\perp}^{-1}$, and energy in units of $\hbar\omega_{\perp}$)
\begin{eqnarray}
i\partial_{t}A_{k} & = & \left[-\dfrac{1}{2}\partial_{x}^{2}+i(-1)^{k-1}\gamma\,\partial_{x}+V(x)\right.\nonumber \\
 &  & \left.+g_{k}\left|A_{k}\right|^{2}+g_{12}\left|A_{3-k}\right|^{2}\right]A_{k}+\Gamma A_{3-k}\,,\label{1D_NLSEs}
\end{eqnarray}
where $A_{k}$ ($k=1,2$) are wave functions related to the two pseudospin
components of the BEC. The strengths of the intra- and interspecies
interactions are $g_{k}\equiv2a_{k}/a_{\perp}$ and $g_{12}\equiv2a_{12}/a_{\perp}$,
with $a_{k}$ and $a_{12}$ being the respective s-wave scattering
lengths. The strengths of the SO and Rabi couplings are $\gamma\equiv k_{L}a_{\perp}$
and $\Gamma\equiv\Omega/(2\omega_{\perp})$, respectively, where $k_{L}$
is the wave number of the Raman lasers that couple the two atomic
hyperfine states in the $x$ direction \cite{Hamner_PRL15}, and $\Omega$
is the frequency of the Raman coupling, responsible for the Rabi mixing
between the states. 

In the following, we will assume a null interspecies interactions
$g_{12}=0$ (which can be properly adjusted by means of the Feshbach
resonance \cite{Inouye_NAT98}), i.e., we consider cases where the
interspecies interaction is provided only by the Rabi term. Also,
in a complete attractive binary BEC (negative $g_{1}=g_{2}=g$ and
$\Gamma$) one can obtain localized solutions even in absence of axial
confinement, because in specific conditions the self-trapping of the
cigar-shaped cloud prevents spreading. In this sense, in our model
we consider $V(x)=0$. In order to investigate the details of this
physical process, specifically in the collisional dynamics of two
solitons, in the next section we derive a reduced ODE model that aims
to provide an effective description of the collision dynamics. 

\section{The reduced ODE model\label{sec:III}}

\begin{figure*}[t]
\begin{centering}
\includegraphics[width=0.85\paperwidth]{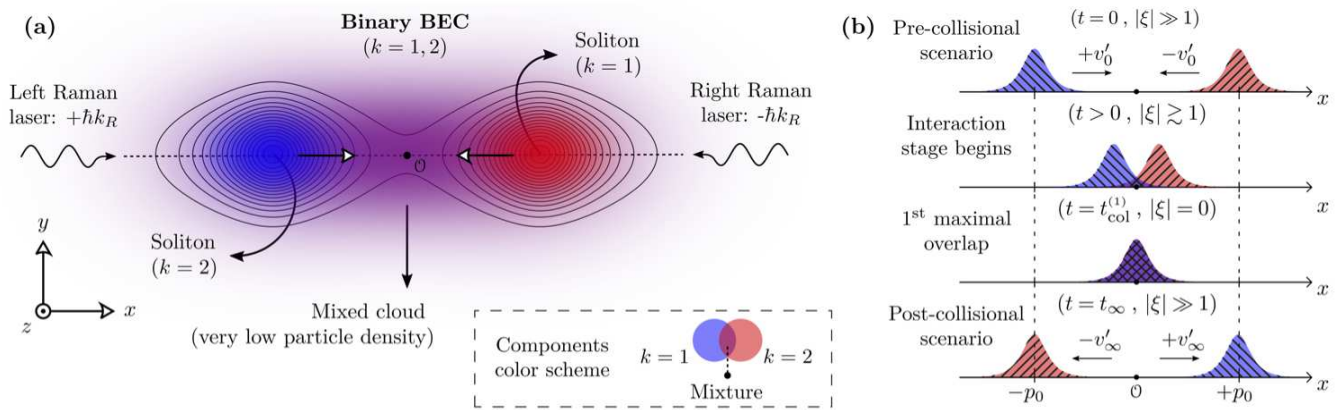}
\par\end{centering}
\caption{(Color online) Pictorial representation of the pre-collisional scenario
of two symmetric solitons in a SO and Rabi-coupled BEC. In (a) and
in the top frame of (b), the pre-collisional scenario consists of
both solitons (initially with peak position at $x=\pm p_{0}$) moving
toward the origin ($\mathcal{O}$) with propagation velocity $\protect\vec{v_{0}}'=(-1)^{k}\left(v_{0}+\gamma\right)\hat{x}$
(for $k=1,2$ and $v_{0}'=v_{0}+\gamma>0$), for the $k$-soliton
component, that is induced by the initial phase velocity and by the
Raman laser field pumped in the $(-1)^{k}\hat{x}$ direction. The
remaining three frames in (b) illustrate the evolution of the initial
configuration, i.e., by showing the beginning of the interaction stage,
which is followed by the first collision process with maximum overlap
at $t=t_{\text{col}}^{\text{\tiny(\ensuremath{1})}}$. The last frame
depicts the post-collisional scenario, with the scattered solitons
moving away from each other with propagation velocity $v_{\infty}'$
and eventually reaching their initial separation at $t=t_{\infty}$.}
\label{Fig1}
\end{figure*}

For convenience, we reset the indexes for the components using the
rule $k\rightarrow\mathrm{sgn}[(-1)^{k}]$ ($k=1,2$). Then, we assume
an approximated solution in a full functional form for symmetric bright
solitons, which can be written in the form
\begin{equation}
A_{\pm}=\eta\,\textrm{sech}\left(\dfrac{x\pm p}{w}\right)\mathrm{e}^{i\left[\pm v\left(x\pm p\right)+\frac{b}{2w}(x\pm p)^{2}+\sigma\right]},\label{ANSATZ}
\end{equation}
with the variational parameters within $A_{\pm}$ being time-dependent
functions, namely: amplitude ($\eta$), velocity ($v$), width ($w$),
peak position ($p$), chirp ($b$), and global phase ($\sigma$).
The exponent comes from the Galilean invariance of Eq. \eqref{1D_NLSEs},
excepting the quadratic term in $x$, which gives a parabolic phase
offset to the waves that promotes width oscillations. The parameter
$\sigma$ develops an important role in the model, because it is responsible
for the global phase invariance of the system. Note that the momentum
conservation arises naturally from the \emph{ansatz}, because the
total momentum of the symmetric solitons is always zero.

The Lagrangian density corresponding to Eq. \eqref{1D_NLSEs} can
be written as $\mathscr{L}=\mathcal{L}_{+}+\mathcal{L}_{-}$, in which
\begin{eqnarray}
\mathcal{L}_{\pm} & = & \Im\left(A_{\pm}^{*}\partial_{t}A_{\pm}\right)\pm\gamma\,\Im\left(A_{\pm}^{*}\partial_{x}A_{\pm}\right)\nonumber \\
 &  & +\dfrac{1}{2}\left|\partial_{x}A_{\pm}\right|^{2}+\dfrac{g_{\pm}}{2}\left|A_{\pm}\right|^{4}+\Gamma\Re\left(A_{\pm}^{*}A_{\mp}\right),\label{LAG_DENS}
\end{eqnarray}
where $\Im(\xi)$ and $\Re(\xi)$ denote the imaginary and real parts
of the complex argument $\xi$, respectively. 

The variational approach yields a reduced ODE model that is calculated
by substituting the \emph{ansatz} \eqref{ANSATZ} into the effective
Lagrangian density \eqref{LAG_DENS}, and then integrating over the
whole $x$-axis. The resulting Lagrangian is given in terms of the
variational parameters and their temporal derivatives, as follows
\begin{align}
L & =4\eta^{2}w\left(v\dot{p}+\dot{\sigma}\right)+\dfrac{\pi^{2}\eta^{2}w}{6}\left(\dot{b}w-b\dot{w}\right)+4\gamma\eta^{2}wv\nonumber \\
 & +2\eta^{2}w\left(v^{2}+\dfrac{1}{3w^{2}}+\frac{\pi^{2}b^{2}}{12}\right)+\dfrac{4g\eta^{4}w}{3}+4\pi\Gamma\eta^{2}wG\ ,\label{RM_LAG}
\end{align}
where the coupling function $G=G\left(\xi,\zeta,w\right)\,$, written
as function of the auxiliary variables $\xi=2p/w$ and $\zeta=2v+\xi b$
plus the parameter $w\,$, is given by
\begin{align}
 & G\left(\xi,\zeta,w\right)=\dfrac{\sin\left(\zeta p\right)}{\sinh\left(\xi\right)\sinh\left(\pi\zeta w/2\right)}\ .\label{G_FUNC}
\end{align}

Since the resulting Lagrangian depends upon the global phase only
through the term $\left(4\eta^{2}w\right)\dot{\sigma\,}$, the Euler-Lagrange
equation for $\sigma$ provides the norm conservation in the reduced
ODE model, i.e., 
\begin{equation}
K=4\eta^{2}w,
\end{equation}
which simply states that $\int_{-\infty}^{\infty}dx\,(\left|A_{+}\right|^{2}+\left|A_{-}\right|^{2})=K\,$,
allowing one to acquire $\eta(t)$ directly from $w(t)\,$. Also,
the other Euler-Lagrange equations arising from the Lagrangian \eqref{RM_LAG}
yield a system of four coupled ODEs, the so-called reduced model,
written as\begin{subequations}
\begin{equation}
\dot{v}=\pi\Gamma\dfrac{\partial G}{\partial p},\label{vp}
\end{equation}
\begin{equation}
\dot{w}=b+\dfrac{12\Gamma}{\pi}\dfrac{\partial G}{\partial b}\label{wp}
\end{equation}
\begin{equation}
\dot{p}=-\left(v'+\pi\Gamma\dfrac{\partial G}{\partial v}\right),\label{pp}
\end{equation}
\begin{equation}
\dot{b}=\dfrac{3}{\pi^{2}}\left(\dfrac{4}{3w^{3}}+\dfrac{gK}{3w^{2}}-4\pi\Gamma\dfrac{\partial G}{\partial w}\right),\label{bp}
\end{equation}
\end{subequations}with $v'=v+\gamma\,$. These equations govern the
evolution of the four independent variational parameters that characterize
the system of symmetric solitons possessing the fixed functional form
given by the \emph{ansatz} \eqref{ANSATZ}.

The set of parameters $\mathcal{C}(t)=\{p(t),v(t),w(t),b(t)\}$ expresses
the configuration of the system at an instant of time $t>0$, which
evolves from an initial configuration $\mathcal{C}_{0}=\{p_{0},v_{0},w_{0},b_{0}\}$
(here we use the notation: $q(0)=q_{0}$). To properly investigate
the scattering of symmetric solitons in this variational model, one
needs to build a set of $\mathcal{C}_{0}$ that corresponds to a desired
pre-collisional scenario. In Fig. \ref{Fig1}, two illustrative representations
of such pre-collisional scenario are shown. In this case, we have
$\left|\xi\right|\gg1$, which means that the separation of the solitons
(given by $2|p|$) is much greater than their width, providing a negligible
tail overlap at the origin of the coordinate system, such that the
system can be represented by two noninteracting symmetric solitons.
This correspondence is no longer valid when the interaction stage
begins, i.e., at the ``moment'' in which the decreasing separation
is $\left|\xi\right|\gtrsim1$, and the increasing overlap of the
solitons' tails eventually becomes large enough so that the effects
of the Rabi interaction becomes substantial. 

We will see (next section) that the interacting solitons can collide
once or several times. In the latter case, they can form a bound-state
that endures until the last collision. Each collision is a process
that mostly affect the dynamics of the solitons during the time near
the instant of maximal overlap (as depicted in Fig. \ref{Fig1}(b)
for the first collision), which is denoted by $t=t_{\text{col}}^{\text{\tiny(\ensuremath{j})}}$
for the $j$-th collision (hence, $p(t_{\text{col}}^{\text{\tiny(\ensuremath{j})}})=0$),
with $j=1\,,\dots,\,n_{\text{col}}$ and $n_{\text{col}}$ being the
total number of collisions during the bound-state. 

These collision processes can induce width oscillations in the solitary
waves. It is a dynamical property that manifests when a part of the
solitons' kinetic energy is contained within a wave profile vibration.
Such property plays a very important role in the bound-state dynamics
and can prevail after the unbinding. So, one can expect that the post-collisional
scenario is characterized by scattered solitons moving away from each
other and endowed with width oscillations (this scenario is illustrated
in Fig. \ref{Fig1}(b) for a transmission case). As their separation
gradually increases, the inequality $\left|\xi\right|\gg1$ eventually
holds, allowing the noninteracting solitons correspondence to be applied
again. 

In this work we focus on the scattering of solitary waves manifesting
in the form of fundamental soliton solutions during the pre-collisional
scenario, this means that the solitons' shape remains practically
the same until the interaction stage (no width change: $\dot{w}=0$).
Width oscillations during the post-collisional scenario are expected
and analytically tractable due to the simplifications allowed\textcolor{blue}{{}
}by the $\left|\xi\right|\gg1$ regime in the reduced model equations
(Eqs. (\ref{vp})-(\ref{bp})). Hence, the width dynamics in this
regime is studied in the next section, which also introduces some
important concepts and definitions regarding the total energy of the
system, which are essential in the discussions concerning the main
issue of this article.

\section{Initial conditions and width oscillations\label{sec:IV}}

In order to build the general form of a set of parameters $\mathcal{C}_{0}$
for pre-collisional scenarios, some basic insight about the solitons'
dynamics in the reduced model is required, and hence the Eqs. \eqref{vp}-\eqref{bp}
need to be analyzed. Firstly, note that in all four equations there
is a term directly proportional to $\Gamma(\partial G/\partial q)$
(with $q=v,\,w,\,p$ or $b\,$), which couples the variational parameters
with each other. When the solitons are far from each other (as in
pre- or post-collisional scenarios), i.e., for $\left|\xi\right|\gg1$,
these coupling terms become negligible since the denominator of $G$
increases very fast for large $|\xi|$ due to a dominating term $\ \propto\exp\left[-\left|\left(1+\pi wb/2\right)\xi\right|\,\right]\,$,
allowing one to assume that $\partial_{q}G\approx0$ and $G\approx0\,$.
In this regime, the reduced model describes noninteracting solitons
(with null acceleration $\dot{v}=0$ in \eqref{vp}) moving toward
(outward) the origin when $\dot{p}<0$ ($\dot{p}>0$), with constant
absolute velocity $\left|v'\right|$ as stated by Eq. \eqref{pp}.
Also, this equation shows that $v$ can be identified as the propagation
velocity (given by $\dot{p}\,$) only in the absence of the SO coupling
($\gamma=0$). Note that the above approximations fails when the solitons
get closer to each other, such that the term $\pi\Gamma\partial_{v}G$
becomes relevant. In fact, the role of the variational parameter $v$
consists in emulating the effect of the phase velocity that, together
with the group velocity $\gamma$ induced by the SO coupling, promote
the collisional scenario of solitons moving initially with propagation
velocity $v_{0}'=v_{0}+\gamma>0$ (as previously pointed out in Fig.
\ref{Fig1}). 

Eqs. \eqref{wp} and \eqref{bp} govern the dynamics of the shape
parameters $(w,b)$. In the regime $\left|\xi\right|\gg1$, the parameter
$b$ dictates the variations in the width, since $b=\dot{w}$, where
the conditions for a fixed profile can be derived by simultaneously
imposing $b=0$ and $\dot{b}=0$. The solutions are $w_{f}=4/\left(\left|g\right|K\right)$
and $b_{f}=0$, with $f$ standing for \textit{fundamental} (without
oscillation). Then, to get a pre-collisional configuration consisting
of fundamental solitons, one can simply use a set of initial parameters
in the form $\mathcal{C}_{0}^{f}=\{p_{0},v_{0},w_{f},b_{f}\}$ such
that $v_{0}>-\gamma$ and $\left|\xi_{0}\right|\gg1$. Next, by considering
slightly different shape parameters, an analytical study of the width
behavior can be directly performed by means of the dynamic equations.
To this end, the width parameter must be rewritten as $w(t)=\left[1+W(t)\right]w_{f}$,
with the new parameter $W(t)\ll1$ being the relative deviation from
$w_{f}$. The latter assumption allows one to expand the Eq. \eqref{bp}
in Taylor series ($(1+W)^{-n}=1-n\,W+{\cal O}(W^{2})$ for $n>0$),
in view to find the following equations: 
\begin{align}
\begin{cases}
w_{f}\dot{W}-b=0\\
W+\dot{b}/\mathcal{B}\hspace{0.5pt}=0
\end{cases} & ,\quad\mathcal{B}=\dfrac{4}{\pi^{2}w_{f}^{3}}\;,\label{LVeqs}
\end{align}
neglecting terms of order ${\cal O}(W^{2})$. The equations (\eqref{LVeqs})
can be cast in a decoupled form $\ddot{q}+\left(\mathcal{B}/w_{f}\right)q=0$
(with $q=W$ or $b\,$), which reveals that both $w$ and $b$ undergo
harmonic oscillations with angular frequency $\omega_{w}^{\textrm{{\tiny\,(LO)}}}=\sqrt{\mathcal{B}/w_{f}}=g^{2}K^{2}/\left(8\pi\right)$
(LO stands for low amplitude oscillations). Additionally, Eqs. \eqref{LVeqs}
show that these parameters oscillate out of phase by $\pi/2$ radians
with oscillation amplitudes $\hat{W}$ and $\hat{b}$ related through
the ratio $\hat{b}/\hat{W}=\omega_{w}^{\textrm{{\tiny\,(LO)}}}$,
hence the condition $\hat{W}\ll1$ implies in $\hat{b}\ll1$.

In the interaction stage, the coupling terms containing $\partial_{q}G$
influence the system's dynamics in a nontrivial way that cannot be
analytically tractable. Since the shape parameters are altered during
the collision processes, width oscillations are expected to occur,
but the behavior is far from being quasi-harmonic because the inequality
$\left|\xi\right|\gg1$ does not hold and $\hat{W}$ is not small.
The latter condition also applies to the post-collisional scenarios,
i.e., the scattered solitons can be provided with highly nonharmonic
width oscillations. To investigate this case, one can explore the
fact that total energy of the system is a conserved quantity, given
by the Hamiltonian
\begin{equation}
H(p,v,w,b)=H_{\textrm{{\tiny TM}}}+H_{\textrm{{\tiny VM}}}+\pi\Gamma\left(G-G_{0}\right),\label{RM_HAM}
\end{equation}
where 
\begin{eqnarray*}
H_{\textrm{{\tiny TM}}}(p,v) & = & \dfrac{1}{2}\left(v+\gamma\right)^{2}+\pi\Gamma G_{0},\\
H_{\textrm{{\tiny VM}}}(w,b) & = & \dfrac{Kg}{12w}+\dfrac{1}{6w^{2}}+\frac{\pi^{2}}{24}b^{2},\\
G_{0}=\left.G\,\right|_{(w,b)=(w_{f},0)} & = & \dfrac{\sin\left(2pv\right)}{\sinh\left(2p/w_{f}\right)\sinh\left(\pi vw_{f}\right)}.
\end{eqnarray*}
The first and the second terms in the Hamiltonian correspond to the
energy within the solitons' translational mode (TM) and vibrational
mode (VM), respectively, and the third is an energy term due to the
interaction of these modes \cite{Yang_10}. The idea of casting the
Hamiltonian as shown in \eqref{RM_HAM} is to highlight the energy
contributions arising from each type of motion of the solitons in
the reduced model.

The Hamiltonian \eqref{RM_HAM} in its entire form will be used in
the next section. For a while, the focus is on the general behavior
of width oscillations emerging in post-collisional scenarios. In this
sense, terms originating from the function $G$ are negligible, allowing
one to identify the solitons' TM energy by their kinetic energy, i.e.,
$H_{\textrm{{\tiny TM}}}(v')=\left(v'\right)^{2}/2$. By considering
the configurations at $t=0$, given by $\mathcal{C}_{0}^{f}$, one
obtains the Hamiltonian
\begin{equation}
H_{0}(v_{0})=H_{\textrm{{\tiny TM}}}^{(0)}+H_{\textrm{{\tiny VM}}}^{(0)},
\end{equation}
where the first term, $H_{\textrm{{\tiny TM}}}^{(0)}=H_{\textrm{{\tiny TM}}}(v'_{0})$,
is the TM initial energy, and the last, $H_{\textrm{{\tiny VM}}}^{(0)}=H_{\textrm{{\tiny VM}}}(w_{0},b_{0})=-g^{2}K^{2}/96\,$,
is the self-energy of the fundamental solitons. After an ``infinitely''
long time interval $\left(t\rightarrow\infty\right)$, the Hamiltonian
of the scattered solitons can be written as
\begin{equation}
H^{(\infty)}=H_{\textrm{{\tiny TM}}}^{(\infty)}+H_{\textrm{{\tiny VM}}}^{(\infty)},\label{RM_HAM_F}
\end{equation}
where $H_{\textrm{{\tiny TM}}}^{(\infty)}=H_{\textrm{{\tiny TM}}}(v_{\infty}')$,
with $v_{\infty}=v(t\rightarrow\infty)$, and
\[
H_{\textrm{{\tiny VM}}}^{(\infty)}=\left.\left(\dfrac{Kg}{12w}+\dfrac{1}{6w^{2}}+\frac{\pi^{2}b^{2}}{24}\right)\right|_{t\rightarrow\infty}.
\]
Here, $v_{\infty}$ is the final (constant) value of the phase velocity,
and $H_{\textrm{{\tiny TM\,(VM)}}}^{(\infty)}$ is the TM (VM) final
energy. We stress that the parameter $v$ approaches $v_{\infty}$
asymptotically during the post-collisional scenario, but in a practical
sense, one can set $t_{\infty}$ as the instant in which the initial
separation is reattained ($p(t_{\infty})=p_{0}$), where $t\rightarrow\infty$
in \eqref{RM_HAM_F} was replaced by $t=t_{\infty}$ (as shown in
the last frame in Fig. \ref{Fig1}(b)). 

The energy conservation implies that $\Delta H=H^{(\infty)}-H^{(0)}=0$.
By using this result combined with the equations $b=\dot{w}$ and
$w=\left(1+W\right)w_{f}$, one can obtain the following equation
for the parameter $W(t)$ ($t>t_{\infty}$) in terms of the initial
and final propagation velocities
\begin{equation}
\left(\pi\,\dot{W}\right)^{2}+\left(\dfrac{g^{4}K^{4}}{64}\right)\dfrac{W^{2}}{(1+W)^{2}}=-\left(\dfrac{3g^{2}K^{2}}{2}\right)\Delta H_{\textrm{{\tiny TM}}}\ ,\label{POST_WO}
\end{equation}
where $\Delta H_{\textrm{{\tiny TM\,(VM)}}}=H_{\textrm{{\tiny TM\,(VM)}}}^{(\infty)}-H_{\textrm{{\tiny TM\,(VM)}}}^{(0)}$
is the TM (VM) energy variation, obeying the relation $\Delta H_{\textrm{{\tiny TM}}}=-\Delta H_{\textrm{{\tiny VM}}}$.
Based on the positiveness of all terms in the left-hand side of the
Eq. \eqref{POST_WO}, the energy variation of the modes are such that
$\Delta H_{\textrm{{\tiny TM}}}\leq0$ and $\Delta H_{\textrm{{\tiny VM}}}\geq0$,
which implies $\left|v_{\infty}'\right|\leq v_{0}'$ (recall that
$v_{0}'>0$) with the equalities holding when the scattered solitons
have no vibrational profile ($W=\dot{W}=0$). Except for this latter
trivial case, $W$ has two critical values (denoted by $W_{c}^{\pm}$)
that are obtained from Eq. \eqref{POST_WO} subjected to the condition
$\dot{W}=0$. These critical values are found to be
\begin{align}
W_{c}^{\pm} & =\pm\dfrac{\sqrt{6\,\Delta H_{\textrm{{\tiny VM}}}\phantom{^{0}}}}{|g|K/4\mp\sqrt{6\,\Delta H_{\textrm{{\tiny VM}}}\phantom{^{0}}}}\ ,\label{CRIT_WIDTH}
\end{align}
with $W_{c}^{+}\geq0$ being the positive critical value and $W_{c}^{-}\leq0$
the negative one. In view to solve the first order differential equation
for $W$, one gets 
\begin{equation}
dt=\dfrac{8\pi}{gK}\dfrac{(1+W)\,dW}{\sqrt{96\,\Delta H_{\textrm{{\tiny VM}}}(1+W)^{2}-g^{2}K^{2}W^{2}\phantom{^{0}}}}\ .\label{W_INTEGRAL}
\end{equation}
Indeed, it appears to be a hard task to solve Eq. \eqref{W_INTEGRAL}
for $W(t)$. However, the behavior of the width parameter is periodic.
So, one can write $t(W_{c}^{+})-t(W_{c}^{-})$ equal to half of the
width oscillation period $\left(\,T_{w}/2\,\right)$. Hence, by using
the relation $\omega_{w}=2\pi/T_{w}\,$, the angular frequency of
width oscillations is found to be
\begin{equation}
\omega_{w}=\omega_{w}^{\textrm{{\tiny\,(LO)}}}\left[1-\dfrac{\Delta H_{\textrm{{\tiny VM}}}}{|H_{\textrm{{\tiny VM}}}^{(0)}|}\right]^{3/2},\hspace*{1em}\Delta H_{\textrm{{\tiny VM}}}\leq|H_{\textrm{{\tiny VM}}}^{(0)}|\,.\label{WIDTH_AFREQ}
\end{equation}

The Eqs. \eqref{CRIT_WIDTH} and \eqref{WIDTH_AFREQ} characterize
the width oscillations in the post-collisional scenario in terms of
the initial and final propagation velocities, $v_{0}'$ and $v_{\infty}'$,
which provide the energy increase in the VM ($\Delta H_{\textrm{{\tiny VM}}}=[\,\left(v_{0}'\right)^{2}-\left(v_{\infty}'\right)^{2}\,]/2\,$).
Since $\left|v_{\infty}'\right|\leq v_{0}'\,$, the scattering can
be of three types, \emph{namely}, elastic (case $|v_{\infty}'|=v_{0}'$),
inelastic (case $|v_{\infty}'|<v_{0}'$), and completely inelastic
(case $|v_{\infty}'|=0$). An elastic scattering occurs when the
TM energy is completely recovered after the interaction stage, resulting
in scattered solitons with fixed shape ($\Delta H_{\textrm{{\tiny VM}}}=0$
and $W_{c}^{\pm}=0$), otherwise the amount of energy not recovered
remains stored in the VM (inelastic scattering), and the scattered
solitons will vibrate ($\Delta H_{\textrm{{\tiny VM}}}>0$ and $W_{c}^{\pm}\neq0$).
If this amount of energy is very small such that $\left|v_{\infty}'\right|\lesssim v_{0}'$
(quasi-elastic scattering), the vibration can be considered to be
quasi-harmonic because $\Delta H_{\textrm{{\tiny VM}}}\ll|H_{\textrm{{\tiny VM}}}^{(0)}|$
implies that $W_{c}^{+}\approx|W_{c}^{-}|\ll1$ and $\omega_{w}\approx\omega_{w}^{\textrm{{\tiny\,(LO)}}}$,
which validate the results of the previous approach for low amplitude
of width oscillations. If the TM final energy is zero, the total energy
of the system is entirely contained in the VM (completely inelastic
scattering, $\Delta H_{\textrm{{\tiny VM}}}=\left(v_{0}'\right)^{2}/2$
or $H^{(\infty)}=H_{\textrm{{\tiny VM}}}^{(\infty)}$), resulting
in scattered solitons with fixed separation vibrating with the largest
(lowest) possible amplitude (frequency). In terms of width oscillations,
this means that for a specific value of $v_{0}'$, the critical values
of $|W|$ are maximum and $\omega_{w}$ is minimum.

Since the knowledge about $v_{\infty}'$ it is enough for us to characterize
both TM and VM dynamics of the scattered solitons, the investigation
of solitons' scattering starts from the choice of initial value of
$v_{0}'$ and its influence over the interaction stage.

\section{Numerical Results and Discussion \label{sec:V}}

We set the value of nonlinearity strength $g$ such that the width
of the fundamental soliton solution is  $w_{f}=1$ and the solitons'
total norm $K=1$. These constraints are attained for $g=-4$. Also,
we set the Rabi coupling as $\Gamma=-0.04$, which allow us to get
interesting dynamical effects. The interaction between the solitons
is sufficiently small for a $20$ units wide separation, which justify
our choice of $p_{0}=10$. The program developed for the simulations
uses double precision for both real and complex numbers, it is written
in the Fortran 95 language and employs a $4^{\text{th}}$-order Runge-Kutta
method to numerically solve the coupled ODEs \eqref{vp}-\eqref{bp}
with initial conditions given by $\mathcal{C}_{0}^{f}(v_{0}')$ and
$v_{0}'=v_{0}+\gamma>0$ being a variable initial parameter defining
the pre-collisional configuration. The time-step is set to $10^{-4}$,
this value is small enough to provide a very good approximation for
the evolution of the variational parameters in the conditions of our
interest. Also, in order to check the accuracy of the results obtained,
we performed some tests by considering lower values of discretization,
for which we obtained similar results.

\begin{figure*}[t]
\begin{centering}
\includegraphics[width=1\textwidth]{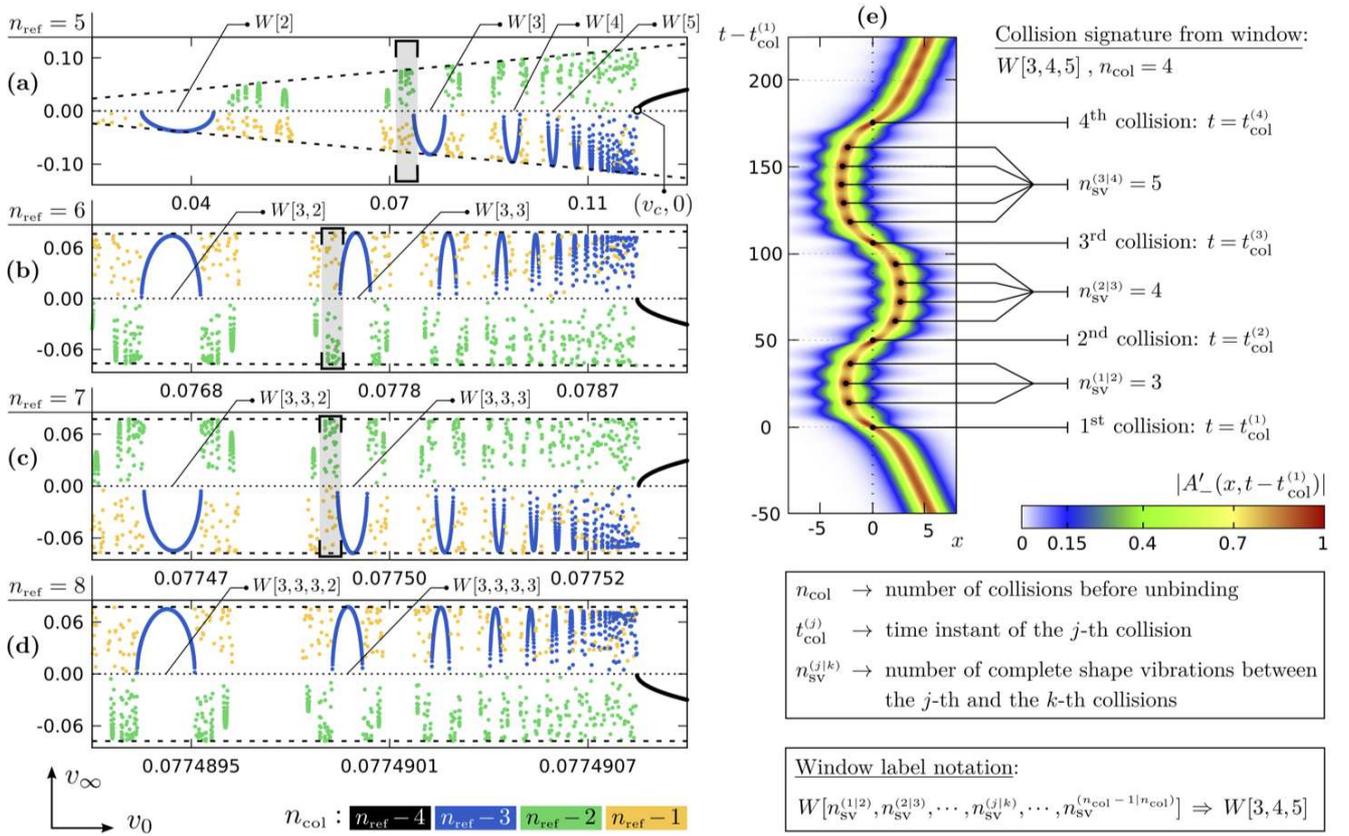}
\par\end{centering}
\caption{(Color online)\textit{ Left panel}: Scattering results for $v_{\infty}$
versus $v_{0}$ obtained via iterative simulations of the reduced
the ODE model (Eqs. \eqref{vp}-\eqref{bp}) in four $v_{0}$-ranges
(a)-(d), within the interval $[0,v_{c}]$ and with $\gamma=0$, i.e.,
without the SO coupling effect. The highlighted rectangular regions
(gray) indicate the $v_{0}$-range of the plot immediately below,
i.e., the panels in (b), (c) and (d) are successive ``zooms'' of
the highlighted regions. The color scheme at the bottom of this figure
uses the integer $n_{\text{ref}}$, called reference number (its value
is specified in top right corner of each plot), to provide an adaptive
rule for coloring the points $(v_{0},v_{\infty})$ accordingly to
the integer $n_{\text{col}}$ associated with the corresponding regular
process (irregular ones are not plotted since $n_{\text{col}}\gg1$).
Also, some windows are labeled in each plot, where the used notations
are explained in the right panel. \textit{Right panel: }(e) heatmap
of the normalized function $|A'_{-}(x,t)|$. The $v_{0}$ value used
in this simulation belongs to the interval of regularity of a $4$-pass
collisional scattering window. The notation used in the description
of this heatmap is explained in the bottom boxes of this panel.}
\label{Fig2}
\end{figure*}

To explore the influence of $v_{0}'$ over the solitons' dynamics,
an iterative routine is implemented to perform a set of consecutive
scattering simulations, each one using a different initial propagation
velocity, $v_{0}'(j)$ (with $j\in\mathbb{N}$ being the iteration
number), which can only assume values within a predefined $v_{0}'$-range
$[v_{0}'(1),v_{0}'(n_{I})]$ (with $n_{I}$ being the total number
of iterations). In this routine, the value of the SO coupling constant
$\gamma$ is kept fixed while $v_{0}$ is increased by a fixed amount
$\delta v_{0}>0$ in the end of each iteration, i.e., $v_{0}'(j+1)=v_{0}'(j)+\delta v_{0}$.
The length of the continuous interval defined by the $v_{0}'$-range
is simply given by the difference between the $v_{0}'$ values used
in the first and in the last scattering simulation, $L=v_{0}'(n_{I})-v_{0}'(1)$,
and consequently $\delta v_{0}=L/(n_{I}-1)$. Moreover, for each scattering
simulation the output data is obtained when the numerical evolution
stops after the program detects that the initial separation was reached
in the post-collisional scenario. In this sense, the quantities analyzed
are the number of collisions before unbinding $n_{\text{col}}$, the
exit velocity $v_{\infty}$, $W_{c}^{+}$, and $T_{w}$. 

We stress that we choose a convenient integer $n_{\text{ref}}$ as
reference and only the points with $n_{\text{ref}}-4\leq n_{\text{col}}\leq n_{\text{ref}}-1$
are considered in our graphical analyzes. So, the remaining points,
with $n_{\text{col}}<n_{\text{ref}}-4$ or $n_{\text{col}}>n_{\text{ref}}$,
are not plotted.

\subsection{Scattering process without SO coupling ($\gamma=0$)}

In this subsection we will consider the system in absence of SO coupling
($\gamma=0$). This first step will provide us a reference for the
dynamical properties, which will be analyzed in details in order to
verify, in the next subsection, the influence of the SO coupling parameter
$\gamma$ over them.

The results of the iterative simulations show that in the high-energy
collision regime ($v_{0}\gg1$) the solitons collide one time ($n_{\text{col}}=1$)
and their phase velocity almost does not diminishes ($v_{\infty}\lesssim v_{0}$),
indicating that the scattering is quasi-elastic and that the single
collision process promotes just a direct transmission (the solitons
simply pass through each other). In this regime, as $v_{0}$ increases
the quantities $v_{\infty}$, $W_{c}^{+}$, and $\omega_{w}$ asymptotically
approach the lines $v_{\infty}=v_{0}$, $W_{c}^{+}=0$, and $\omega_{w}=\omega_{w}^{\textrm{{\tiny\,(LO)}}}=2/\pi$,
respectively, which are associated with the ``scattering'' of two
noninteracting symmetric solitons. As $v_{0}$ is reduced, the scattering
gradually becomes more inelastic, that is, $\Delta H_{\textrm{{\tiny VM}}}$
increases causing $W_{c}^{+}$ to increase too and $\omega_{w}$ to
decrease. When $v_{0}$ is close to the value $v_{\textrm{{\tiny(VM)}}}=0.374$,
the excitation of the vibrational mode is maximum although the variation
in the translational mode energy is still relatively small (since
$v_{\infty}\approx0.898\,v_{0}$), this means that $W_{c}^{+}$ is
maximum too and $\omega_{w}$ is minimum, with $\max_{G}(W_{c}^{+})\approx0.397$
and $\min_{G}(\omega_{w})\approx0.881\,\omega_{w}^{\textrm{{\tiny\,(LO)}}}$
(the estimates were obtained from graphical analyses, and $G$ stands
for global, i.e., for any $v_{0}>0$). Accordingly, as $v_{0}$ gets
even smaller (low-energy collision regime $v_{0}<v_{\textrm{{\tiny(VM)}}}$),
$\Delta H_{\textrm{{\tiny VM}}}$ and, consequently, $W_{c}^{+}$
decreases too (the opposite stands for $\omega_{w}$). The origin
of this inversion in the behavior of these quantities can be understood
by analyzing the equation $\Delta H_{\textrm{{\tiny VM}}}=(v_{0}-v_{\infty})(v_{0}+v_{\infty})/2$
for decreasing $v_{0}$. The first factor always grows because the
scattering becomes more inelastic, and it dominates during the high-energy
collision regime. On the contrary, the second factor always declines
due the reducing amount of energy involved in the first collision,
it exactly balances the growth promoted by the first one when $v_{0}=v_{\textrm{{\tiny(VM)}}}$,
and dominates during the low-energy collision regime causing $\Delta H_{\textrm{{\tiny VM}}}$
to decrease. This behavior persists until $v_{0}$ reaches a critical
value $v_{c}\approx0.11755$, that corresponds to a completely inelastic
scattering ($v_{\infty}=0$). If $v_{0}<v_{c}$, the solitons form
a bound-state after the first collision process and $n_{\text{col}}\geq2$.
The scattering simulations in this range reveal that the dynamics
of this bound-state is very complex and rich in details, requiring
a quite extensive investigation in order to understand the underlying
mechanism produced by the attractive Rabi interaction. From hereafter,
the focus is on the correlations between the output quantities and
the control (input) parameter $v_{0}'\in(0,v_{c}')$, and how these
arise from the reduced model description of the solitons' bound-state.

In Fig. \ref{Fig2}, the left panel shows four plots of $v_{\infty}\times v_{0}$,
which were generated from the data provided by the iterative simulations.
Specifically, the panel (a) (with $n_{\text{ref}}=5$) covers a $v_{0}$-range
in the low-energy collision regime, where $(v_{c},0)$ can be seen
as a critical-point that separates the region of direct transmission,
or 1-pass collisional scattering (points with $n_{\text{col}}=n_{\text{ref}}-4=1$,
see the color scheme at the bottom of the figure), from the region
of multi-pass collisional scattering, where the post-collisional scenario
is always preceded by the formation of a bound-state (points with
$n_{\text{col}}=2,3$ and $4$). The distribution of points in this
plot reveals that $v_{\infty}$ and $n_{\text{col}}$ obey the equation
$\text{sign}(v_{\infty})=(-1)^{n_{\text{col}}-1}$, which states that
a transmission-like scattering ($v_{\infty}>0$) always occurs when
$n_{\text{col}}$ is odd, and a reflection-like scattering ($v_{\infty}<0$)
always occurs when $n_{\text{col}}$ is even.

Regarding the region of 1-pass collisional scattering (Fig. \ref{Fig2}(a)),
one can verify that the points $(v_{0},v_{\infty})$ closely trace
the upper segment of a hyperbola with functional form $x^{m}-y^{m}=v_{c}^{m}$
(with $v_{0}$ and $v_{\infty}$ taking the roles of $x$ and $y$,
respectively), which has its vertex in the critical point and its
asymptotes (the lines $y=\pm x$) represented by the dashed lines
in Fig. \ref{Fig2}. Next, by a fitting procedure we get $m=1.814\pm0.007$,
showing that the collision outcome can be predicted very accurately
when $v_{0}\geq v_{c}$. This control is possible because a small
variation in the initial velocity $v_{0}\rightarrow v_{0}+\delta$,
causes a small variation in its final velocity $v_{\infty}\rightarrow v_{\infty}+\Delta$,
with $\delta$ and $\Delta$ having the same order of magnitude, and
the scattering is said to be regular in this sense. On the other hand,
the same does not hold when $v_{0}<v_{c}$, since $v_{\infty}$ is
found to be very sensitive to small changes in the values $v_{0}$
for some regions. Indeed, there are some regions with regularity for
$v_{0}<v_{c}$, in which we can obtain predictable results. The most
evident intervals of regularity are those where only 2-pass collision
scattering ($n_{\text{col}}=2$) happens, called reflection windows,
which are seen as valley-like shapes in Fig. \ref{Fig2}(a). The asymptote
$y=-x$ is tangent to the curve defined by all these shapes, which
means that an elastic 2-pass collision scattering is possible for
a specific $v_{0}$ value within each reflection window. Interestingly,
these windows appear to form a structure that presents self-similarity
at any scale (a fractal-like scattering), i.e., any amplification
of a smaller $v_{0}$-range containing the critical point reveals
(given an enough point density) the same pattern of infinitely many
reflection windows intertwined by regions in which $n_{\text{col}}>2$.
This happens because both the length of a window and its separation
distance to the nearest window can become arbitrarily small as close
as it gets to the critical point.

Regarding 3-pass collisional scattering ($n_{\text{col}}=3$), the
Fig. \ref{Fig2}(a) shows that it can happen if the $v_{0}$ value
is sufficiently close to one of the edges of any reflection window,
where some of the associated points are found to be within very small
intervals of regularity, which technically requires a much higher
local point density to be reasonably visualized. Therefore, in order
to verify how these points are really distributed, iterative simulations
were performed in $v_{0}$-ranges near the left and right sides of
certain reflection windows. The complementary data acquired unfolds
some substructures of transmission windows that were previously hard
to detect, and strongly indicate that 3-pass collisional scatterings
can only occur when $v_{0}$ falls into an interval of regularity
corresponding to one of these transmission windows, assuming lump-like
shapes in Figs. \ref{Fig2}(b). These substructures are endowed with
the same self-similarity property previously discussed, but only those
emerging at the left side of a reflection window present a pattern
that resembles the one shown in the panel (a) (the windows height
in right-sided substructures decrease instead of increasing accordingly
with the asymptote $y=x$). Indeed, the range that encompasses the
larger left-sided substructure, highlighted by a rectangular (gray)
region in Fig. \ref{Fig2}(a), was simulated again with more points
and displayed in Fig. \ref{Fig2}(b). This plot provides a wide view
of the particular substructure chosen in Fig. \ref{Fig2}(a), where
one can notice that both the window pattern and the distribution of
points near the windows edges are indeed very similar (``mirrored'')
to that of the first plot. 

By investigating the surroundings of the transmission windows through
some iterative simulations, other smaller substructures associated
with 4-pass collisional scatterings are revealed. These are composed
by reflection windows too and present a high degree of similarity
with the previous plot, a signature of the fractal-like scattering,
as one can attest by comparing it with the plot in Fig. \ref{Fig2}(c),
which considers the left-sided substructure of the second transmission
window (the highlighted (grey) region in Fig. \ref{Fig2}(b)). Thus,
all the plotted points within the intervals intertwining the reflection
windows in the panel (a) are part of underlying substructures, which
unfold whenever one investigates the distribution of points surrounding
any reflection or transmission window.

The whole structure composed by infinitely many reflection and transmission
windows displays the main characteristic feature of a fractal, i.e.,
self-similarity. Here, such fractal-like consists of the main window
pattern ($n_{\text{col}}=2$) plus the left-sided (right-sided) ones
associated with $n_{\text{col}}\,$-pass collisional scatterings ($n_{\text{col}}\geq3$)
that emerge in subregions within $v_{0}<v_{c}\,\land\,|v_{\infty}|\leq v_{0}$
that contain only the left (right) critical (or edge) point of a certain
$(n_{\text{col}}-1)$-pass collisional scattering window, which is
a point corresponding to a completely inelastic $(n_{\text{col}}-1)$-pass
collisional scattering. A much higher degree of self-similarity is
clearly noticed between the window patterns of the substructures,
as one can realize by comparing Fig. \ref{Fig2}(b) and (c), which
appear to be mirrored images (across the $v_{0}$-axis) from each
other. The plot displayed in Fig. \ref{Fig2}(d) results from iterative
simulations in the range highlighted in Fig. \ref{Fig2}(c), it emphasizes
the fractal feature described and show that the window pattern replicates
more precisely in substructures that have the same type of window.
As previously mentioned, another feature regarding the solitons' scattering
is its high sensitivity to $v_{0}$ when this initial propagation
velocity is not within an interval of regularity, this is a signature
of chaos that allows us to infer that the scattering is predominantly
chaotic when $v_{0}<v_{c}$, which is intrinsically related to the
formation of bound-states generally involving a lot of collisions
(i.e., $n_{\text{col}}\gg1$, excepting the region of very low propagation
velocities at the left of the larger reflection window). Hence, the
fractal structure must arise from a recurrent internal mechanism that
causes the scattering to become regular when specific conditions involving
the solitons' translational and vibrational modes are attained. We
stress that the fractal scattering of solitons of systems described
by (generalized) nonlinear Schr\"odinger equation were also verified
in Refs. \cite{Zhu_PRL08,Zhu_PD08,Dmitriev_CHAOS02,Teixeira_PLA16}.

To unravel this internal mechanism, a detailed analysis of the solitons'
dynamics during the interaction stage is needed. To this end, we first
study the general aspects of the bound-states by examining the evolution
of the solitons' profile from the perspective of the heatmaps of $|A'_{-}(x,t)|$.
For that, several simulations are performed for different values of
$v_{0}$ selected in some intervals of those reflection and transmission
windows shown in Fig. \ref{Fig2}(a)-(d). By analyzing the bound-state
formation for various input velocities within a same interval, one
can only differ one scattering from another by comparing the shape
vibrations and the exit angle ($\tan^{-1}(v_{\infty})$) in the post-collisional
scenario, that is, before the final collision the dynamics is visually
indistinguishable (this is more prominent when considering smaller
windows). This means that each window has its own bound-state signature
describing the consistent behavior of the solitons' modes that gives
rise to the window itself. Moreover, this signature is unique and
can be simply defined in terms of the number of complete shape vibrations
(a full width oscillation period) between two consecutive collisions
during the bound-state, as indicated in Fig. \ref{Fig2}(e). This
full width oscillation period is taken as a time interval centered
in an instant $t=t_{\text{peak}}$ of minimum profile width (or maximum
profile amplitude). In this way one can count the number of peaks
(spots in the heatmap where $|A'_{-}(x,t)|\apprle1$) between the
$(j-1)$-th and the $j$-th collisions ($j=2,\dots,n_{\text{col}}$)
and assign the resulting integer value to $\ensuremath{n_{\text{sv}}^{\text{\tiny(\ensuremath{j-1|j})}}}$
(see the notation introduced in Fig. \ref{Fig2}). Then, any $n_{\text{col}}\,$-pass
collisional scattering window can be labeled in terms of these $n_{\text{col}}-1$
integers as pointed out by Fig. \ref{Fig2}(e), where the heatmap
displayed corresponds to the 4-pass collisional scattering window
$W[3,4,5]$.

Interestingly, the window signatures also follow a pattern that is
naturally connected with the fractal-like structure. It is first seen
in the panel (a), where the label of the $j$-th window (always from
left to right) is written as $W[j+1]$, i.e., $n_{\text{sv}}^{\text{\tiny(1\ensuremath{|}2)}}=j+1$.
Then, based on the consistent window patterns previously discussed,
one can infer from the heatmaps analysis that the $k$-th window of
the substructure emerging from the left side of $W[j+1]$ can be labeled
as $W[j+1,k+1]$ (see Fig. \ref{Fig2}(b)), with the changing index
$J=k+1$ defined as the main index. The same applies for the $l$-th
window of the substructure emerging from the left side of $W[j+1,k+1]$,
which has the label $W[j+1,k+1,l+1]$ ($J=l+1$ is the main index
here), and so on.

The integers $\ensuremath{n_{\text{sv}}^{\text{\tiny(\ensuremath{j-1|j})}}}$
($j=2,\dots,n_{\text{col}}$) that define a $n_{\text{col}}\,$-pass
collisional scattering window signature depend on the frequency of
the shape vibration ($\ensuremath{\omega_{\text{sv}}}$) and on the
time duration of each bounce $\Delta t_{\text{\,bounce}}^{\text{\tiny\,(\ensuremath{j-1|j})}}=t_{\text{col}}^{\text{\tiny(\ensuremath{j})}}-t_{\text{col}}^{\text{\tiny(\ensuremath{j-1})}}$.
In analyzing the shape parameters evolution, we verified that $\ensuremath{\omega_{\text{sv}}}$
is approximately constant during the bouncing time intervals between
collisions, when the tail overlap is small enough so that the interaction
promotes an effective attraction maintaining the solitons' bound-state
while exerting a weak influence over the previously induced shape
oscillations. Also, we found that the quantity $\Delta t_{\text{\,bounce}}^{\text{\tiny\,(\ensuremath{n_{\text{col}}-1|n_{\text{col}}})}}$
(time duration of the last bounce) strictly increases with $v_{0}$
as it covers the entire interval (from left to right) of a $n_{\text{col}}\,$-pass
collisional scattering window (sub)structure, with the corresponding
critical point being a singularity in which $\Delta t_{\text{\,bounce}}^{\text{\tiny\,(\ensuremath{n_{\text{col}}-1|n_{\text{col}}})}}\rightarrow\infty$. 

If $v_{0}$ is within the interval of regularity of a window with
main index $J$, i.e., $\ensuremath{n_{\text{sv}}^{\text{\tiny(\ensuremath{n_{\text{col}}-1|n_{\text{col}}})}}=J}$,
one can write $\Delta t_{\text{\,bounce}}^{\text{\tiny\,(\ensuremath{n_{\text{col}}-1|n_{\text{col}}})}}=J\ensuremath{T_{\text{sv}}}+\delta t_{\text{col}}$,
in which $\ensuremath{T_{\text{sv}}}=2\pi/\ensuremath{\omega_{\text{sv}}}$
is the shape vibration period and $\delta t_{\text{col}}$ is a $v_{0}$
dependent term accounting for the time duration associated with the
$(n_{\text{col}}-1)$-th and $n_{\text{col}}$-th collisions when
$\ensuremath{\omega_{\text{sv}}}$ is no longer constant. We found
that this linear behavior for $\Delta t_{\text{\,bounce}}^{\text{\tiny\,(\ensuremath{n_{\text{col}}-1|n_{\text{col}}})}}$
as function of $J$ occurs when the left or the right edge points
of five consecutive windows ($J=1,\dots,5$) are considered. In this
case, $\delta t_{\text{col}}$ tends to assume the same value when
$v_{0}$ is about to leave the intervals of regularity. The angular
coefficient of the fitting line provides a reasonable estimate of
$\ensuremath{T_{\text{sv}}}$, which was obtained with standard deviation
always less than $2\%$ for two sets of five points of each plot in
Fig. \ref{Fig2}. Concerning the structure in the panel (a), the average
value obtained was $\left\langle \ensuremath{T_{\text{sv}}}\right\rangle =10.8\pm0.2\ (1,8\text{\%})$,
while for the substructures in the panels (b)-(d) the average values
of $\ensuremath{T_{\text{sv}}}$ are the same, given by $\left\langle \ensuremath{T_{\text{sv}}}\right\rangle =9.98\pm0.02\ (0.21\text{\%})$.
The numerical quantity $2\pi/\left\langle \ensuremath{T_{\text{sv}}}\right\rangle \approx0.63$
is\textcolor{blue}{{} }a reasonable estimate for the shape vibration
frequency,\textcolor{blue}{{} }which indicates that such vibrational
motion in regular processes have indeed a characteristic frequency.

Next, we analyze the behavior of $\delta t_{\text{col}}$ in terms
of $v_{0}$. We found that this quantity strictly increases with $v_{0}$
such that $1.7\lesssim\delta t_{\text{col}}/\ensuremath{T_{\text{sv}}}\lesssim2.0$,
with the left (right) sided extreme value reached when $v_{0}$ assumes
the value corresponding to the left (right) edge of a window. So,
it follows that the bouncing frequency $\omega_{\text{\,bounce}}^{\text{\tiny\,(\ensuremath{n_{\text{col}}-1|n_{\text{col}}})}}=2\pi/\Delta t_{\text{\,bounce}}^{\text{\tiny\,(\ensuremath{n_{\text{col}}-1|n_{\text{col}}})}}$
must approximately satisfy the relation
\begin{equation}
\left(J+1.85+d\right)\omega_{\text{\,bounce}}^{\text{\tiny\,(\ensuremath{n_{\text{col}}-1|n_{\text{col}}})}}=\omega_{\text{sv}}\quad\left(\ \left|d\right|\lesssim0.15\ \right),\label{BounceSV}
\end{equation}
which establishes the condition of motion synchronization involving
the solitons' translational and vibrational modes, which give rise
to the intervals of regularity. This condition means that the bouncing
motion is such that $\omega_{\text{\,bounce}}^{\text{\tiny\,(\ensuremath{n_{\text{col}}-1|n_{\text{col}}})}}$
must approach a state of resonance with the shape vibration, $\omega_{\text{\,bounce}}^{\text{\tiny\,(\ensuremath{n_{\text{col}}-1|n_{\text{col}}})}}=\omega_{\text{sv}}/(J+3)$,
from below by a suitable amount provided by Eq. \eqref{BounceSV}.
The process is always irregular if $\omega_{\text{\,bounce}}^{\text{\tiny\,(\ensuremath{n_{\text{col}}-1|n_{\text{col}}})}}$
is not close enough to or exceeds a resonance value ($\omega_{\text{sv}}/3,\,\omega_{\text{sv}}/4,\,\omega_{\text{sv}}/5,\,\dots$).
The narrowing of the windows of a given structure results from the
behavior of $\omega_{\text{\,bounce}}^{\text{\tiny\,(\ensuremath{n_{\text{col}}-1|n_{\text{col}}})}}$
with $v_{0}$, which decreases faster as close as $v_{0}$ is from
the corresponding critical value in a such way that the greater the
integer $J$ is, smaller is the $v_{0}$-interval in which the condition
\eqref{BounceSV} holds and, consequently, narrower is the window.

\begin{figure}[t]
\begin{centering}
\includegraphics[width=1\columnwidth]{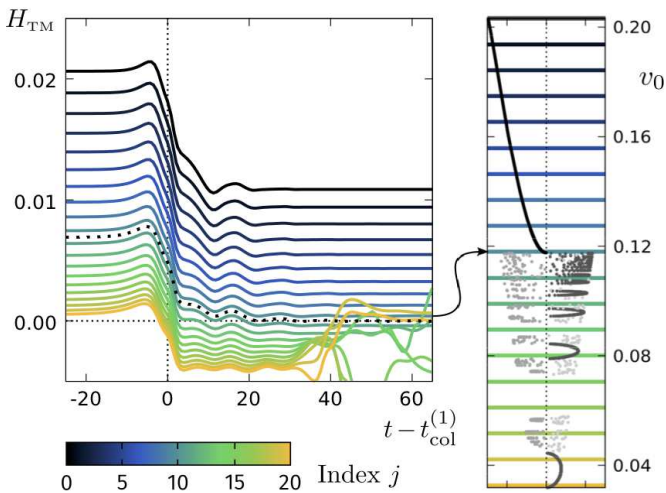}
\par\end{centering}
\caption{(Color online) Energy of the solitons' translational mode ($H_{\textrm{{\tiny TM}}}$)
as function of the time variable $t-\ensuremath{t_{\text{col}}^{(1)}}$,
by considering $20$ scattering processes (indexed with integers $j\in[1,20]$
in the heatmap). The $v_{0}$ values for each one of these processes
is highlighted by horizontal lines crossing the rotated version of
the plot seen in Fig. \ref{Fig2}(a) (right side). The dashed vertical
line at the time $t=\ensuremath{t_{\text{col}}^{(1)}}$ highlights
the instant of the first collision, i.e., when the interaction causes
an effective decrease in $H_{\textrm{{\tiny TM}}}$, resulting in
part of the initial TM energy converted in VM energy, which in turn
promotes shape vibrations. For the $10$ last processes ($j=11,\dots,20$),
$v_{0}<v_{c}$ and $H_{\textrm{{\tiny TM}}}$ becomes negative right
after the $t=\ensuremath{t_{\text{col}}^{(1)}}$.}
\label{Fig3}
\end{figure}

\begin{figure*}[t]
\begin{centering}
\includegraphics[width=0.85\paperwidth]{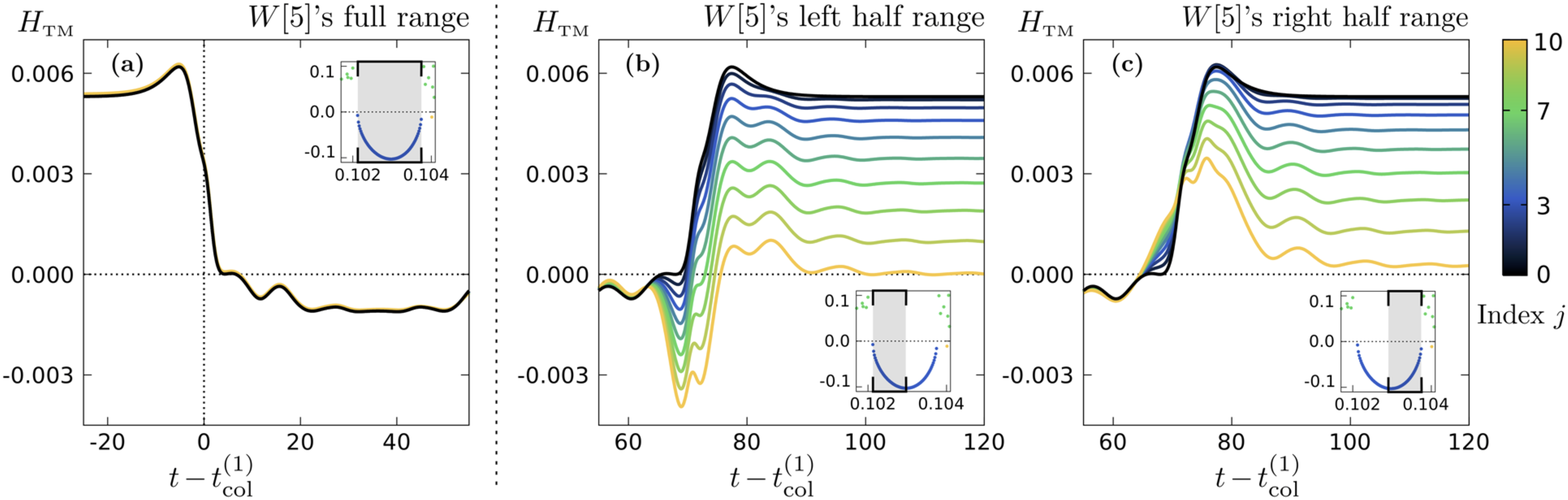}
\par\end{centering}
\caption{(Color online) Energy of the solitons' translational mode ($H_{\textrm{{\tiny TM}}}$)
as function of the time variable $t-\ensuremath{t_{\text{col}}^{(1)}}$,
by considering $20$ scattering processes. Here, the $v_{0}$ values
for each one of these processes are taken in the $W[5]$'s interval
of regularity analogously to the plot in Fig. \ref{Fig3}. In the
panel (a), $v_{0}$ covers the full range (as indicated by the inset,
containing the corresponding $v_{\infty}\times v_{0}$ plot). In this
case, the $20$ energy plots are almost indistinguishable. In the
panels (b) and (c), $v_{0}$ covers the left and the right half range,
respectively, starting from the middle point and then toward the edges
(see the inset panels). In each plot, $10$ processes are displayed
and indexed with integers $j\in[1,10]$. The time range starts from
the final point shown in panel (a). In $t-\ensuremath{t_{\text{col}}^{(1)}}=120$,
the solitons will have spread out and $H_{\textrm{{\tiny TM}}}\approx H_{\textrm{{\tiny TM}}}^{(\infty)}$.}
\label{Fig4}
\end{figure*}

From section \ref{sec:IV}, we bring back the quantities defined in
\eqref{RM_HAM_F} to investigate the scattering mechanism in terms
of the energy within the solitons' modes. To this end, we firstly
considered $20$ distinct scattering processes with $v_{0}$ varying
into the interval $[v_{c}-\Delta v,v_{c}+\Delta v]$ from right to
left, with $v_{c}-\Delta v$ chosen to match the $v_{0}$ value of
$W[2]$'s left edge point. The right half of this interval is in the
direct transmission region, i.e., the first $10$ processes are regular
ones consisting of just one collision. In Fig. \ref{Fig3}, the temporal
evolution of $H_{\textrm{{\tiny TM}}}$ is shown for each scattering
process, with the index $j\in[1,20]$. We observe that for $j\in[1,10]$
($v_{0}>v_{c}$), the first collision effectively causes a decrease
in the energy of the TM that reaches a stable positive constant value
($H_{\textrm{{\tiny TM}}}^{(\infty)}=\text{const.}>0$) as the solitons
get far apart from\textcolor{blue}{{} }each other. For $j=10$, the
associated $v_{0}$ value is very close to $v_{c}$ and $H_{\textrm{{\tiny TM}}}^{(\infty)}\apprge0$.
This is an expected result since the ``critical process'' ($v_{0}=v_{c}$)
must end up with $H_{\textrm{{\tiny TM}}}^{(\infty)}=0$.

From Fig. \ref{Fig3}, we verify that for $j\in[11,20]$ ($v_{0}<v_{c}$),
$H_{\textrm{{\tiny TM}}}$ is negative and oscillatory (sometimes
reaching the positive range again) until the moment of the last collision
($t=t_{\text{\,col}}^{\text{\tiny\,(\ensuremath{n_{\text{col}}})}}$,
which is close to the time $t-t_{\text{\,col}}^{\text{\tiny\,(\ensuremath{n_{\text{col}}})}}=40$
for the two last processes with $v_{0}$ within $W[2]$). In this
case the TM recovers enough energy to remain positive (unbinding)
and eventually constant as the separation between the solitons increases.
Therefore, these results show that a final negative TM energy value
is a signature of the formation of bound-states. Also, during the
evolution of this state one can attest that $H_{\textrm{{\tiny TM}}}$
is indeed a predominantly negative valued function of time, i.e.,
it can eventually becomes positive valued for a short time without
triggering the unbinding and then return to the negative range, but
we attested that this can happen only in chaotic processes. Into the
regular windows, when $H_{\textrm{{\tiny TM}}}$ oscillates and reach
the positive range, the solitons unbind and scatter away ($H_{\textrm{{\tiny TM}}}\rightarrow H_{\textrm{{\tiny TM}}}^{(\infty)}>0$). 

To clarify the above statement, we proceeded as before by considering
$20$ distinct scattering processes with $v_{0}$ now covering a full
window range. In Fig. \ref{Fig4}, the energy of the solitons' TM
($H_{\textrm{{\tiny TM}}}$) are shown for $v_{0}$ into the $W[5]$'s
interval of regularity. In \ref{Fig4}(a) we observe that, before
the last (second) collision, $H_{\textrm{{\tiny TM}}}$ is not affected
by changes in $v_{0}$. This is because all variational parameter
display this same behavior embedded in $H_{\textrm{{\tiny TM\,(VM)}}}$,
which prevails until the second collision, for which subtle differences
accumulated during the bound-state evolution become enough to promote
very different interaction outcomes, as one can note in Figs. \ref{Fig4}(b)
and \ref{Fig4}(c). In fact, based on extensive analyses of the simulations
data, we were able to infer that this initial dynamics of the modes
energy is maintained until the eminence of the last collision for
all observed collection of scattering processes within an arbitrary
window $W[\dots,J]$. Also, it extends similarly for any irregular
process in the chaotic region nearby, i.e., if the condition of motion
synchronization (\ref{BounceSV}) is not met, the solitons do not
unbind and any variation in $v_{0}$ causes the upcoming bound-state
dynamics to radically diverge, giving rise to the $v_{\infty}$'s
great sensitivity to $v_{0}$.

\begin{figure}[tp]
\begin{centering}
\includegraphics[width=1\columnwidth]{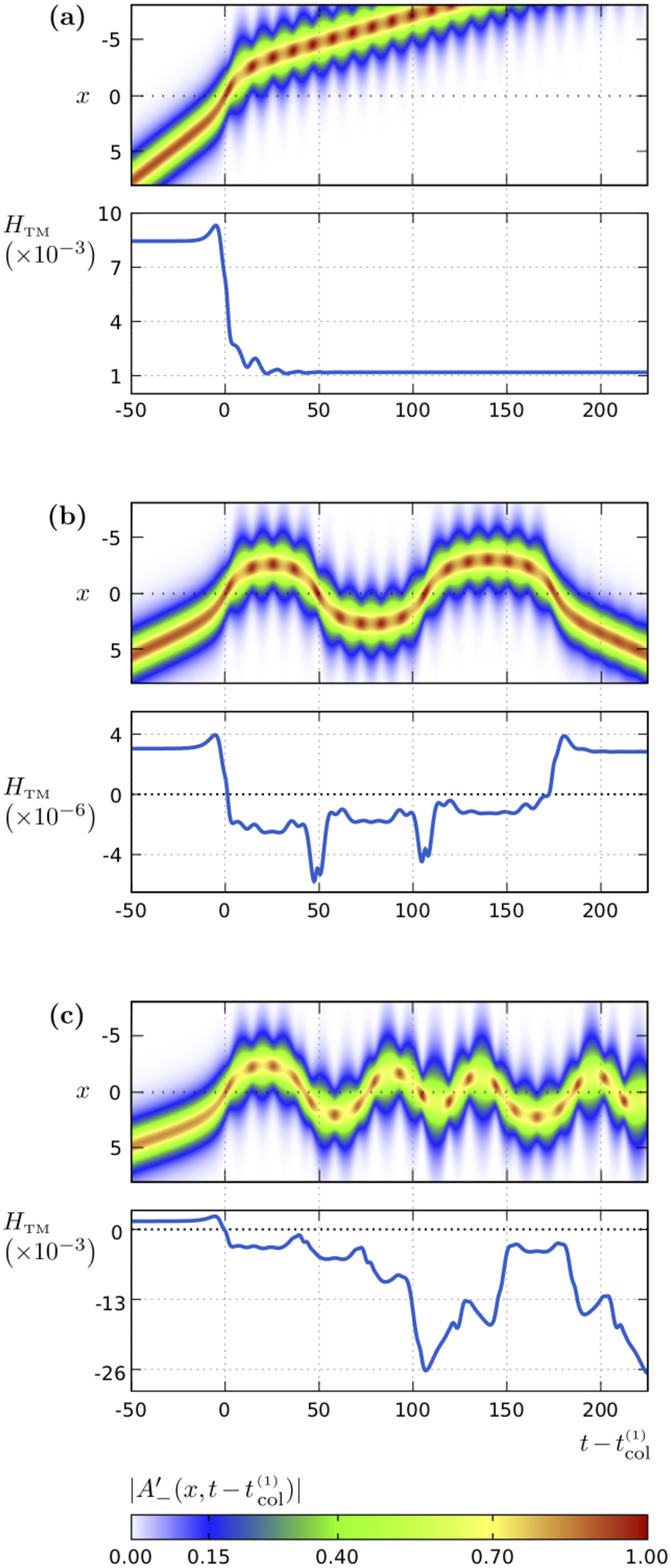}
\par\end{centering}
\caption{(Color online) Heatmap of $|A'_{+}(x,t-\ensuremath{t_{\text{col}}^{(1)}})|$
and the corresponding TM energy ($H_{\textrm{{\tiny TM}}}$) versus
$t-\ensuremath{t_{\text{col}}^{(1)}}$, for three examples of scattering:
(a) direct transmission, (b) regular scattering, and (c) irregular
scattering. In panel (a) it was used $v_{0}=0.13>v_{c}$. In panel
(b) it was set $v_{0}=0.077984$, belonging to $W[3,4,5]$'s interval
of regularity. In panel (c) it was considered $v_{0}=0.055$, which
is located in a chaotic interval between the windows $W[2]$ and $W[3]$.}
\label{Fig5}
\end{figure}

Following, in Fig. \ref{Fig5} we display the profile $|A'_{+}(x,t-\ensuremath{t_{\text{col}}^{(1)}})|$
and the corresponding $H_{\textrm{{\tiny TM}}}$ versus $t-\ensuremath{t_{\text{col}}^{(1)}}$
in order to clarify the basic features regarding both the bound-state
and the TM energy dynamics, for each one of the three cases considered
in this plot. Note that the heatmap in Fig. \ref{Fig5}(a) shows that
the collision induce shape vibrations, as also indicated by the corresponding
$H_{\textrm{{\tiny TM}}}$ evolution, where one can see that the TM
energy is always positive and $H_{\textrm{{\tiny TM}}}^{(\infty)}<H_{\textrm{{\tiny TM}}}^{(0)}$,
as expected since part of the initial energy is transferred to the
VM. Also, the heatmap shown in Fig. \ref{Fig5}(b) is an example of
a regular scattering, as previously displayed in Fig. \ref{Fig2}(e).
This example of regular  process is useful for illustrating that the
longer the bounce time duration ($\Delta t_{\text{\,bounce}}^{\text{\tiny\,(\ensuremath{j-1|j})}}$)
is, smaller is the absolute value of the TM energy. Indeed, this process
occurs because the solitons weakly bind to each other during these
well behaved bounces, due to their separation. On the other hand,
in irregular processes the bound-state frequently evolves to situations
in which the solitons strongly bind to each other, which are characterized
by very high bouncing frequencies $\omega_{\text{\,bounce}}^{\text{\tiny\,(\ensuremath{j-1|j})}}$
(or collision rates) that maintain the average separation very small.
The heatmap from the example in Fig. \ref{Fig5}(c) illustrates such
behavior. It takes place just after the second collision and is accompanied
by a large effective decrease in $H_{\textrm{{\tiny TM}}}$, which
reaches a range of negative values that are greater than $H_{\textrm{{\tiny TM}}}^{(0)}$
by more than an order of magnitude (in modulus). In fact, one can
infer about the binding strength by testing the inequality $|H_{\textrm{{\tiny TM}}}|\gg|H_{\textrm{{\tiny TM}}}^{(0)}|$,
and then infer about the type of scattering process.

\subsection{Effects of SO coupling in the scattering process ($\gamma\protect\neq0$)}

In the previous subsection, we considered the reduced ODE model in
the absence of SO-coupling ($\gamma=0$), where the results of several
scattering simulations revealed the existence of a very rich and complex
dynamics that emerges when the initial velocity is smaller than a
certain threshold value (i.e., $v_{0}<v_{c}$). Our extensive analysis
of the data allowed us to better understand the underlying mechanism
that gives rise to the many interesting features of the solitons in
the variational description. Now we explore what happens with all
these features when the SO-coupling is present ($\gamma\neq0$).

\begin{figure*}[!t]
\begin{centering}
\includegraphics[width=0.75\paperwidth]{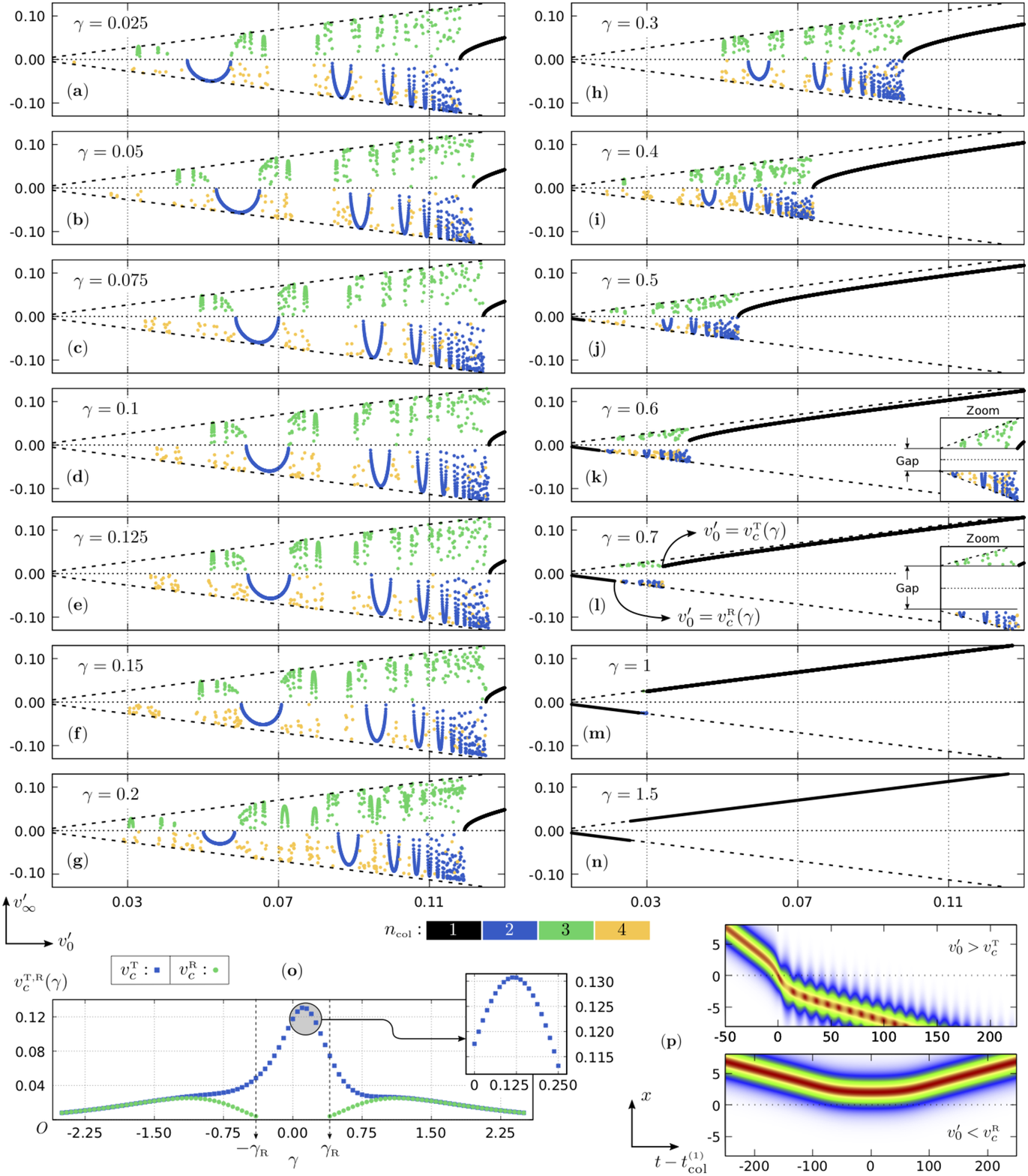}
\par\end{centering}
\caption{(Color online) Scattering results for $v_{\infty}'$ versus $v_{0}'$
obtained via iterative simulations of the reduced ODE model (Eqs.
\eqref{vp}-\eqref{bp}) in a fixed $v_{0}$'-range for different
values of $\gamma>0$, starting from $\gamma=0.025$ in panel (a)
and adding $\Delta\gamma=0.025$ at each step until $\gamma=0.15$
in panel (f). Next, we start from $\gamma=0.2$ in panel (g) and adding
$\Delta\gamma=0.1$ at each step until $\gamma=0.7$ in panel (l).
The SO-coupling parameter is chosen as $\gamma=1.0$ and $\gamma=1.5$
in panels (m) and (n), respectively. At the right corner of the plots
in (k) and (l), a zoom of the window structure is displayed to highlight
the emerging gap that splits the chaotic-like region in two parts.
In panel (o), the approximate values for the two types of critical
velocity, $v_{c}^{\text{\tiny R}}(\gamma)$ and $v_{c}^{\text{\tiny T}}(\gamma)$,
are shown in two graphs, with the smallest one focusing in the $\gamma$-range
where $v_{c}^{\text{\tiny T}}(\gamma)$ reaches a peak value. In panel
(p), two heatmap plots are displayed to exemplify the characteristic
dynamics of the two types of direct scattering, namely, direct reflection
($v_{0}'\leq v_{c}^{\text{\tiny R}}(\gamma)$) and direct transmission
($v_{0}'\geq v_{c}^{\text{\tiny T}}(\gamma)$). }

\label{Fig7}
\end{figure*}

In section \ref{sec:IV}, we have previously pointed out that the
initial propagation velocity cannot be identified with the parameter
$v_{0}$ when $\gamma\neq0$, instead it is $v_{0}'=v_{0}+\gamma$
as indicated by Eq. (\ref{pp}) in the regime $\left|\xi\right|\gg1$.
Regarding only the effective soliton dynamics, as can be seen in heatmap
plots, a pre-collisional scenario with $v_{0}=V_{0}$ and $\gamma=0$
is indistinguishable from one with $v_{0}=V_{0}-\gamma$ and $\gamma\neq0$,
since $v_{0}'=V_{0}$ in both cases. Hence, in order to simulate the
effects of the SO-coupling over pre-collisional scenarios equivalent
to those from the previous subsection, we have used a $v_{0}$-range
similar to that from Fig. \ref{Fig2}(a) translated by $\gamma$ units
to the left (right) if $\gamma>0$ ($\gamma<0$). In Fig. \ref{Fig7},
the effect of the SO-coupling over the final propagation velocity
$v_{\infty}'$ is shown for several cases in which $\gamma>0$. The
plots in Fig. \ref{Fig7}(a)-(g) display similar window structures
that basically differ from another one by some sort of transformation
combining translation and scaling of the intervals of regularity.
The critical point that separates the chaotic-like region from the
regular one also translates along the $v_{0}'$-axis as the $\gamma$
increases. One can realize that $v_{c}$ grows from Fig. \ref{Fig7}(a)
to \ref{Fig7}(d) and diminishes from Fig. \ref{Fig7}(d) to \ref{Fig7}(g).
Besides these changes in the windows placement, there are new transmission
windows associated with 3-pass collisional scattering processes that
now appear at left side of $W[2]$.

We emphasize that for $0<\gamma\lesssim0.2$ the effect of the SO-coupling
in the variational dynamics is small, in the sense that it does not
affect significantly the main structure of windows and its substructures.
So, the mechanism described in the previous subsection still works
when the SO-coupling is present and, after some analysis of the collision
dynamics within several intervals of regularity, one can verify that
those interesting features associated with the reflection/transmission
windows remain. We performed several iterative simulations considering
$\gamma$ values gradually increasing from $0.2$ up to $1.5$ with
step $\Delta\gamma=0.05$. By comparing the obtained plots (some of
which are shown in Fig. \ref{Fig7}(h)-(n)), one can notice that the
critical velocity keeps decreasing as $\gamma$ increases, causing
the whole window structure to be displaced toward the origin. Indeed,
the window closest to the origin shrinks and eventually disappears
when $\gamma$ reaches a certain value. The beginning of this process
can be seen in the window $W[2]$ (left to right) in Fig. \ref{Fig7}(g).
As this process goes on, the structure ``loses'' some windows and
becomes smaller.\textcolor{blue}{{} }When $\gamma=0.5$ (see Fig. \ref{Fig7}(j)),
the structure can be barely seen and becomes even more confined due
to the emergence of a new type of critical point that separates the
chaotic-like region from a new one that extends until the origin ($v_{0}'=0$).
This new region increases with $\gamma$ and speeds up the vanishing
process of the chaotic-like region and the window structures within
it, which are lastly seen in Fig. \ref{Fig7}(l). Following, in Fig.
\ref{Fig7}(m) the window structure is gone, and only a few points
can be barely seen within what is left from the chaotic-like region,
which has already completely vanished in Fig. \ref{Fig7}(m). Comparing
these last two figures, we verify an inversion of the initial increasing
behavior of the new region, since its interval was shortened. 

Back to Fig. \ref{Fig7}(l), we introduce a notation to differ the
new type of critical velocity from the old one, with $v_{c}^{\text{\tiny R}}(\gamma)$
denoting the former and $v_{c}^{\text{\tiny T}}(\gamma)$ the latter
(previously denoted by $v_{c}$). Here the dependence with $\gamma$
is written explicitly, and the superscripts R and T stand for reflection
and transmission, respectively. With this notation we mean that every
scattering process with $v_{0}'>v_{c}^{\text{\tiny R}}(\gamma)$ is
a direct transmission, and that every scattering process with $0<v_{0}'<v_{c}^{\text{\tiny R}}(\gamma)$
is a direct reflection. The latter is a new type of regular scattering
that cannot occur if $\gamma$ does not exceed a certain threshold
value $\gamma_{R}$. As an example, in Fig. \ref{Fig7}(p) it is considered
two plots displaying the behavior of these two types of direct scattering
process. In the direct reflection scenario (bottom plot of Fig. \ref{Fig7}(p))
one can note that the peak position $p$ never reaches zero (without
passing) and that there is no detectable shape vibrations after the
collision, i.e., the scattering is practically elastic (the corresponding
points in the $v_{\infty}^{\prime}\times v_{0}^{\prime}$\textcolor{red}{{}
}plots closely trace the line $y=-x\ |\ x\in[0,v_{c}^{\text{\tiny R}}(\gamma)]$,
as can be seen in Fig. \ref{Fig7}(j)-(n)). Defining $v_{c}^{\text{\tiny R}}(\gamma)=0\ \forall\ \gamma\ |\ 0\leq\gamma<\gamma_{\text{\tiny R}}\ $,
then the direct reflection critical point ($P_{R}$) always coincides
with the origin of the coordinates system (i.e., $P_{R}=(0,0)$),
and the direct transmission one is simply $P_{T}=(v_{c}^{\text{\tiny T}}(\gamma),0)$
as usual. For $\gamma>\gamma_{R}$, the results allows one to write,
in a general way, that $P_{R}\approx(v_{c}^{\text{\tiny R}}(\gamma),-v_{c}^{\text{\tiny R}}(\gamma))$
and that $P_{T}=(v_{c}^{\text{\tiny T}}(\gamma),V_{\infty}^{\text{\tiny T}}(\gamma))$,
with the exit velocity function defined as $V_{\infty}^{\text{\tiny T}}(\gamma)=f(\gamma)v_{c}^{\text{\tiny T}}(\gamma)$,
such that $f(\gamma)=\Theta(\gamma-\gamma_{\text{\tiny R}})\,r_{\gamma}$,
with $\Theta$ being the Heaviside step function and $r_{\gamma}\in[0,1]$.
By graphically tracking the $P_{T}$ point, we found that $r_{\gamma}$
strictly increases with $\gamma$ and asymptotically approaches the
value $1$, as shown in Fig. \ref{Fig7}(n) where $r_{\gamma}\approx1$,
so that $P_{T}$ is very close to the line $y=x$. This means that
scattering process associated with this critical point tends to become
elastic one, with solitons simply crossing each other with almost
no excitation of the vibrational mode.

In order to check the behavior of the critical points $P_{T}$ and
$P_{R}$ with more accuracy, i.e. for a smaller $\Delta\gamma$, we
developed a numerical algorithm to locate these points within a precision
$\log_{10}(\delta v_{c})\leq-5$ and without performing long iterative
simulations over wide $v_{0}'$-ranges. We set $\Delta\gamma=0.05$
and executed the algorithm for $\gamma$ values into the interval
$[-2.5,2.5]$. The corresponding results are shown in Fig. \ref{Fig7}(o).
We found that the $P_{R}$ points distribution is symmetric with respect
to the $\gamma=0$ axis, and also that none of these appear in the
interval $[-\gamma_{\text{\tiny R}},\gamma_{\text{\tiny R}}]$ (as
indicated by our previous analysis for $\gamma>0$). Then, we can
extend the $f$ function to the negative domain by redefining it as
$f(\gamma)=\Theta(|\gamma-\gamma_{\text{\tiny R}}|)\,r_{\gamma}$,
with $r_{\gamma}\approx1$ for SO-coupling strengths $|\gamma|\gg1$.
Additionally, the length of the direct reflection region is maximum,
$\max\left[v_{c}^{\text{\tiny R}}(\gamma)\right]$, when $|\gamma|$
is about $1.15$, and, $v_{c}^{\text{\tiny R}}(|\gamma|)$ strictly
decreases for greater SO-coupling strengths. Regarding the $P_{T}$
points distribution, we observe that it is not symmetric and displays
a special behavior when $\gamma\in[0,\gamma_{\text{\tiny R}}]$. In
this interval, one notes that, for a certain SO-coupling strength
$\gamma_{\text{\tiny T}}>0$, the length of the chaotic-like region
is maximum (i.e., $\max\left[v_{c}^{\text{\tiny T}}(\gamma)\right]=v_{c}^{\text{\tiny T}}(\gamma_{\text{\tiny T}})$).
By reducing the discretization to $\Delta\gamma=0.0125$ over the
interval $[0,0.25]$ (highlighted by an arrow in Fig. \ref{Fig7}(o)),
we obtained that $\gamma_{\text{\tiny T}}$ is about $0.1125$. Indeed,
this result was expected since such behavior could be inferred from
our previous analysis for $\gamma>0$. The asymmetric behavior of
$v_{c}^{\text{\tiny T}}(\gamma)$ displayed in Fig. \ref{Fig7}(o)
is explained as follows. For a SO-coupling strength $|\gamma'|$,
there are always two initial phases giving the same initial propagation
velocity $V_{0}$, which are $v_{0}^{\pm}=V_{0}\pm|\gamma'|$ for
$\gamma=\mp|\gamma'|$. The first term in Eq. (\ref{pp}) is simply
$-v'$, hence it is equal to $-V_{0}$ for both initial conditions
$v_{0}^{\pm}$. Now, if the dependence of the coupling function $G$
with variational parameter $v$ was through a term proportional to
$v'$, then the reduced model would be clearly symmetric with respect
to $\gamma$. However, this is not the case here, because the Rabi
coupling has broken the SO-coupling inversion symmetry. 

Regarding the rest of the $P_{T}$ distribution points residing in
the intervals $[-2.5,-\gamma_{\text{\tiny R}}]$ and $[\gamma_{\text{\tiny R}},2.5]$,
the data shows that $v_{c}^{\text{\tiny T}}(\gamma)$ strictly decreases
for increasing $|\gamma|$. Also, from Fig. \ref{Fig7}(o), we observe
that when $|\gamma|\gtrsim\max[v_{c}^{\text{\tiny R}}]$ the difference
given by $v_{c}^{\text{\tiny T}}(|\gamma|)-v_{c}^{\text{\tiny R}}(|\gamma|)$
(length of the chaotic-like region) is of the order of $10^{-4}$
and quickly approaches $0^{+}$ as $|\gamma|$ grows, i.e., the $P_{T}$
and $P_{R}$ points tend to coalesce for large values of the SO-coupling
strength. In the regime $|\gamma|\gg1$, one can infer from the behavior
of the critical points that $v_{c}^{\text{\tiny T,R}}(|\gamma|)\approx0$,
therefore the scattering tends to become a simple elastic direct transmission
for any pre-collisional scenario ($\forall\,v_{0}'>0$), which is
equivalent to turning off the Rabi coupling.

\begin{figure}[tb]
\begin{centering}
\includegraphics[width=1\columnwidth]{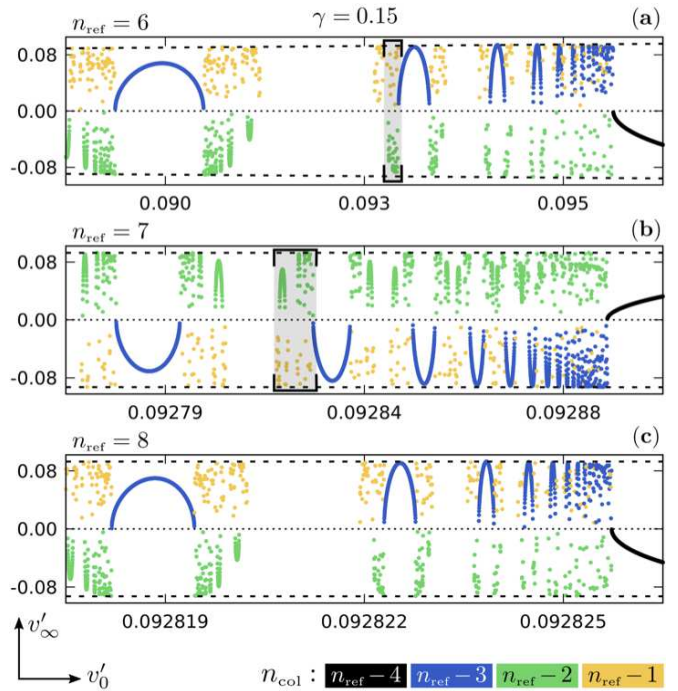}
\par\end{centering}
\caption{(Color online) Scattering results for $v_{\infty}'$ versus $v_{0}'$
obtained via iterative simulations of the reduced ODE model (Eqs.
\eqref{vp}-\eqref{bp}) in three $v_{0}$'-ranges (a)-(c) within
the interval $(0,v_{c}^{\text{\tiny T}}(\gamma)]$ with $\gamma=0.15$.
The two highlighted regions (gray) indicate the $v_{0}$'-range of
the plot immediately below. The panel (a) corresponds to a ``zoom''
of the $v_{0}'$-range containing a substructure near the left edge
of the second window (from left to right) of the main structure displayed
in Fig. \ref{Fig7}(f).}
\label{Fig8}
\end{figure}

By following the same protocol employed in the previous subsection,
we considered here the cases in which $\gamma=\pm0.15$ and investigated
some substructures. As example, in Figs. \ref{Fig8})(a)-(c) we display
the case with $\gamma=0.15$ (similar results are found for the case
with negative sign). When analyzing the window distributions, we found
that the pattern associated with reflection windows differs from the
one associated with transmission windows, with the former having an
overall larger window spacing compared with the latter. However, for
the case $\gamma=-0.15$ one finds an opposite behavior. Hence, the
results indicate that the fractal-like behavior can indeed persist
if the first window structure is weakly affected by the SO-coupling,
and that the changes in the window patterns depend of the sign of
$\gamma$. We also explored some cases in which the SO-coupling strength
caused the chaotic-like regions to become very small as in Figs. \ref{Fig7}(j)-(k).
So, we found that the first substructures still emerge in the edges
of the remaining windows that were not significantly affected by the
vanishing process previously described. 

Hitherto, we have focused on the emergent effects caused by the SO-coupling,
hence our analyses considered only the general aspects regarding the
two types of regular scattering and their associated intervals, with
more emphasis in the intertwining chaotic-like interval and window
structures within it. We have firstly investigated how the parameter
$\gamma$ modifies the coupling function $G$ and its derivatives
$\partial_{q}G$. To this end, we rewrite the Eq. (\ref{G_FUNC})
in terms of the propagation velocity by making $v=v'-\gamma$, which
is equivalent to the variable exchange $\zeta\,\rightarrow\,\zeta'-2\gamma$,
with $\zeta'=2v'+\xi b$ being analogously to $\zeta$ in the case
of $\gamma=0$. Then, defining $G'$ given by
\begin{align}
G'(\xi,\zeta',w) & =\dfrac{\sin\left(\zeta'p-2\gamma p\right)}{\sinh\left(\xi\right)\sinh\left(\pi\zeta'w/2-\pi\gamma w\right)}\ ,\label{G'_FUNC}\\
\left.G'\right|_{b=0,w=1} & =\dfrac{\sin\left[2p(v'-\gamma)\right]}{\sinh\left(2p\right)\sinh\left[\pi(v'-\gamma)\right]}\ \ \left(\text{\small|\ensuremath{\xi}|\,\ensuremath{\gg}\,1}\right),\label{G'_b0}
\end{align}
with the Eq. (\ref{G'_b0}) valid before the collision. We stress
that in the case of $\gamma=0$, the Eq. (\ref{G'_FUNC}) recovers
the form of $G$ obtained in the previous section, i.e., $[G',\zeta',v']{}_{\gamma=0}=[G,\zeta,v]$
(see Eq. (\ref{G_FUNC})). We performed an extensive study of the
above functions to figure out how the terms $2\gamma p$ and $\pi\gamma w$
modify the variational dynamics, with focus on the derivatives $\partial_{p}G'$
and $\partial_{v}G'$, which are associated with the translational
acceleration terms in the reduced model and develop a more important
role in the propagation dynamics. By this study we retrieved the most
important qualitative aspects of the SO-coupling influence over the
interaction.

Considering only the denominator of Eq. (\ref{G'_FUNC}), the term
$\pi\gamma w$ alters the interaction strength in different ways depending
on the behavior of the width parameter $w$. During the bound-states,
the oscillatory character of $w$ due shape vibrations induces oscillations
in the Rabi interaction strength, which are small when $\pi|\gamma|w\ll1$,
i.e., if $|\gamma|\ll1$. For greater SO-coupling strengths, this
oscillation can make the bound-state dynamics very complicated, as
the binding strength keeping the solitons together alternates between
weak and strong regimes. When the SO-coupling is such that $|\gamma|\gg v'$,
the leading effect of the term $\pi\gamma w$ is the dumping of the
Rabi interaction strength, as one can clearly verify from Eq. (\ref{G'_b0}).
This can be related to the behavior of the critical point $P_{T}$,
because, as the Rabi interaction weakens due to the increasing $|\gamma|$,
the maximum propagation velocity for the bound-state formation ($v_{c}^{\text{\tiny T}}(\gamma)$)
reduces until a certain value in which the attraction is still enough
to trap the solitons. On the other hand, regarding now the numerators
of Eqs. (\ref{G'_FUNC}) and (\ref{G'_b0}), one notes that the parameter
$\gamma$ induces oscillations that develop when the solitons are
moving, which occur at a fixed frequency $2|\gamma|$ when $p$ varies
linearly during pre-collisional scenarios. This leads to oscillations
in the sign of every term containing a derivative of $G'$, causing
the Rabi interaction to oscillate between regimes of attraction ($\Gamma\partial_{p,v}G'<0$)
and repulsion ($\Gamma\partial_{p,v}G'>0$). Since the denominator
of Eq. (\ref{G'_b0}) is dominated by the term $\sinh(2p)\gg1$, the
approximation $G',\partial_{q}G'\approx0$ is valid and the sign of
the coupling terms do not matter during pre-collisional scenarios.
Therefore, the sign oscillation become relevant only when $p$ is
small enough so that the translational acceleration terms, $\Gamma\partial_{p,v}G'$,
can significantly alter the propagation. During the bound-states,
$p$ is confined to a narrow interval of values ($|p|\lesssim5$),
if the SO-coupling strength is small, such that $2|\gamma p|\ll1$,
then the sign oscillation barely alters the predominantly attractive
Rabi interaction. In contrast, for greater SO-coupling strengths,
$2|\gamma p|$ is not small and such oscillations are much more prominent,
making the oscillations of the bound-state to vary in an unpredictable
way. For instance, one of the consequences of this non trivial behavior
is displayed in Figs. \ref{Fig7}(k)-(l), where one can see a gap
in the chaotic-like region that splits it into two smaller regions,
i.e., there is a forbidden range of final velocities that establishes
a threshold value for $|v_{\infty}'|$ if $v_{0}'\in[v_{c}^{\text{\tiny R}}(\gamma),v_{c}^{\text{\tiny T}}(\gamma)]$.
This effect happens because the Rabi interaction becomes momentarily
repulsive just after the unbinding, and then, due the proximity of
the solitons, the acceleration is greater enough to increase the propagation
velocity. The gain in velocity is greater as greater the SO-coupling
strengths is and also when the acceleration acts for a longer time,
i.e., if $v'$ is very small just after the unbinding (as in those
regular inelastic processes near the window edges). This increasing
gap explains the behavior of the parameter $r_{\gamma}$ in the $P_{T}$
critical point expression, since $V_{\infty}^{\text{\tiny T}}(\gamma)$
follows the gap upper boundary.

\begin{figure*}[t]
\begin{centering}
\includegraphics[width=0.85\paperwidth]{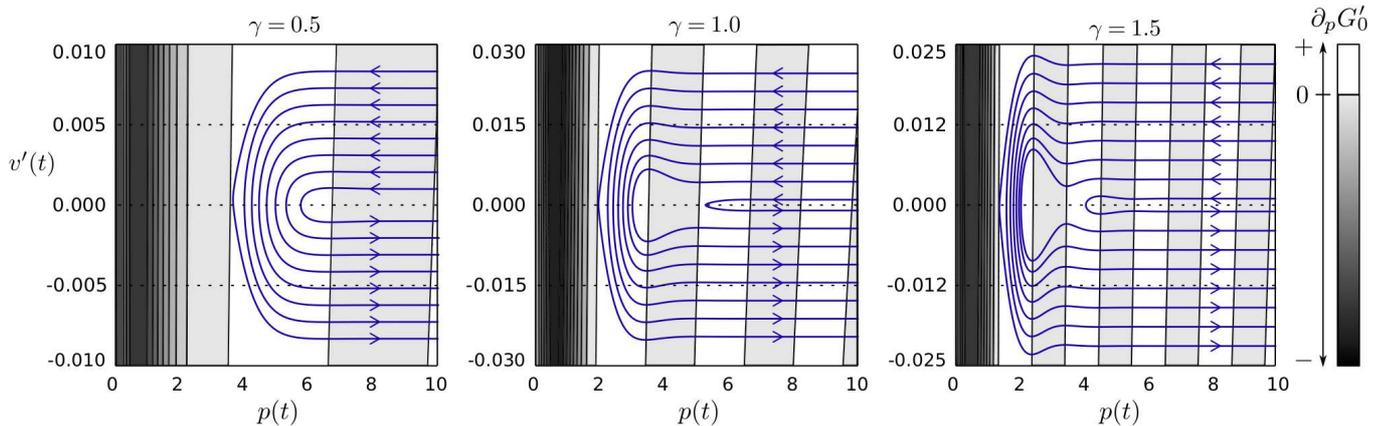}
\par\end{centering}
\caption{(Color online) Phase space trajectories governed by the effective
reduced ODE model (given by\textcolor{blue}{{} }Eqs. (\ref{vp_model}))
providing a variational description for the direct reflection type
of solitons scattering. In the three cases considered ($\gamma=0.5$,
$1.0$, $1.5$), a total of $8$ trajectories with $p_{0}=10$ and
$v_{0}'\in(0,v_{c}^{\text{\tiny R}}(\gamma)]$ are plotted. The background
is a contour line plot of the function $\partial_{p}G_{0}'.$ Since
$\Gamma=-0.04<0$, the attraction zones are the highlighted by gray/black
regions, corresponding to $\text{sgn}\left(\partial_{p}G_{0}'\right)=-1$,
while repulsion zones are identified by the white regions, corresponding
to $\text{sgn}\left(\partial_{p}G_{0}'\right)=+1$.}
\label{Fig9}
\end{figure*}

The alternation between attractive and repulsive Rabi interaction
can be directly related with the emergence of the direct reflection
region. We investigated several scattering processes with $v_{0}'\in(0,v_{c}^{\text{\tiny R}}(\gamma)]$
for various values of $\gamma$, observing that the role of the variational
parameters $w$ and $b$ is negligible. Indeed, this occurs because
the collision is quasi-elastic, with the energy stored within the
vibrational mode being practically zero when compared with the energy
within the translational mode. This finding allowed us to study this
type of scattering in a more quantitative way, since we can set $w=1$
and $b=0$ to obtain the effective reduced ODE model
\begin{equation}
\dot{v}=\pi\Gamma\dfrac{\partial G_{0}'}{\partial p}\quad,\quad\dot{p}=-\left(v'+\pi\Gamma\dfrac{\partial G_{0}'}{\partial v}\right)\ ,\label{vp_model}
\end{equation}
with $G_{0}'=\left.G'\right|_{b=0,w=1}$ being the effective coupling
function yielded by Eq. (\ref{G'_b0}). Note that if $\gamma=0$,
$G_{0}^{\prime}$ coincides with $G_{0}$ of Eq. (\ref{RM_HAM}),
introduced in section \ref{sec:IV}. We study the phase space trajectories
governed by Eq. (\ref{vp_model}) subjected to the initial conditions
$(p_{0},v_{0}')$, with $p_{0}=10$ as usual and $v_{0}'(i)=V_{0}^{\text{(init})}+\Delta_{\gamma}(i-1)\ |\ i\in[1,2,\dots,8]$,
with $\Delta_{\gamma}=(v_{c}^{\text{\tiny R}}(\gamma)-V_{0}^{\text{(init})})/(I-1)$
and $V_{0}^{\text{(init})}=0.001$. When these phase space trajectories
(two-dimensional curves) are plotted with the contour line plot of
$\partial_{p}G_{0}'$ or $\partial_{v}G_{0}'$ in the background,
one can visualize how the propagation is driven by the oscillatory
Rabi interaction, and also how the SO-coupling strength increases
the frequency of such oscillations and consequently alters the dynamics.
This is exactly what is displayed in Fig. \ref{Fig9} for three different
values of $\gamma>0$ and with background composed by the contour
line plots of $\partial_{p}G_{0}'$. The corresponding negative values
provide the same results and similar plots are obtained when $\partial_{v}G_{0}'$
is considered instead. The alternation between attraction (gray zones
with $\Gamma\partial_{p}G_{0}'>0$) and repulsion (white zones with
$\Gamma\partial_{p}G_{0}'<0$) is clearly depicted in Fig. \ref{Fig9}.
Considering the case with $\gamma=0.5$, the trajectories show that
the attraction zone immediately affecting all the processes in the
far field (close to $p=10$) is negligible ($\partial_{p,v}G_{0}'\approx0$)
due to the initially large separation. As $p$ reduces and reaches
the repulsion zone ($p\simeq6$), the separation becomes small enough
to cause a deceleration that can act during a long enough time interval
to completely break the solitons ($v'=0$), and then accelerate them
away ($v'<0$) back to the far field in such way that, in the post-collisional
scenario, $v_{\infty}'\approx-v_{0}'$. Also, one can see that the
shortest trajectory ($v_{0}'=0.001$) quickly turns back as it gets
into the repulsion zone, and that the longest trajectory ($v_{0}'\approx v_{c}^{\text{\tiny R}}(\gamma)$)
turns back after almost reaching the attraction zone that extends
all the way toward $p=0$. 

Regarding the other two cases, with $\gamma=1.0$ and $\gamma=1.5$,
an analogous behavior can be visualized. However, due to the greater
SO-coupling strengths, there is more zones of attraction and repulsion
that add more details to the dynamics. In both cases, the effect of
the attraction/repulsion zones in the far field are once again negligible,
and most of trajectories begin to be significantly affected after
reaching the next-to-last repulsion zone, which is the zone where
the shortest trajectory turns back before reaching the last and most
effective repulsion one (see Fig. \ref{Fig9}). In the case with $\gamma=1.0$,
one observes that the next-to-last repulsion zone barely influences
the other trajectories ($v_{0}'>0.001$), which make the way through
the attraction zone until finally reaching the last repulsion one
and then turning back. In addition, in the case with $\gamma=1.5$,
these final zones are narrower and closer to the $p=0$ axis, hence
the acceleration effects are amplified causing the trajectories to
assume the shapes as seen in Fig. \ref{Fig9}. For greater SO-coupling
strengths, the zones depicted in this figure keep getting narrower
and closer to the $p=0$ axis. The effectiveness of the acceleration
and deceleration under the trajectories diminishes and the maximum
velocity for the occurrence of direct reflection scattering becomes
smaller (this connects with the decreasing behavior of $v_{c}^{\text{\tiny R}}(|\gamma|)$
for $|\gamma|\gtrsim1.15$). As $|\gamma|$ increases further, the
effects of the repulsion and attraction zones cancel each other out
(in average). Also, in this case the Rabi interaction is weakened,
i.e., the scattering tends to be a mere direct transmission for almost
all $v_{0}'>0$.

\section{Conclusion \label{sec:Conclusion}}

In summary, we investigated the influence of the SO coupling on the
collisional dynamics of solitons in binary BECs by using a reduced
ordinary differential equations (ODE) model based on a variational
approach, which allow us to analytically investigate the formation
of fractal-like\textcolor{blue}{{} }patterns and the properties of the
scattered solitons. To this end, we first studied the collision of
solitons in the absence of SO coupling and then we started to verify
the influence on the scattering patterns by changing the value of
the SO coupling parameter $\gamma$. We found exotic structures of
scattering by focusing on the values of the exit velocities $v_{\infty}^{\prime}$
for given input velocities $v_{0}^{\prime}$. Also, we verified that
these structures present a fractal-like pattern, i.e., periodic repetitions
of the main structure in its substructures, corresponding to the zoomed
views. The size of the region presenting windows structures is drastically
affected by the SO coupling. Indeed, we observe that for $|\gamma|\gtrsim1.15$
the structure of windows vanishes completely. Also, the SO-coupling
promotes non-trivial oscillations in the Rabi interaction strength
and its sign, which are the sources of the emergent effects altering
the window structure that vanishes as the chaotic-like region is compressed
in the $v_{0}^{\prime}$-direction by the regions of direct transmission
and direct reflection, and in the $v_{\infty}^{\prime}$-direction
by the growing gap of forbidden final propagation velocities. 

\subsection*{Acknowledgments}

We acknowledge financial support from the Brazilian agencies CNPq,
CAPES, FAPEG, and the National Institute of Science and Technology
(INCT) for Quantum Information.

\bibliographystyle{apsrev4-1}
\phantomsection\addcontentsline{toc}{section}{\refname}\bibliography{Refs}

\end{document}